\documentclass[a4paper,11pt]{article}
\pdfoutput=1 
\usepackage[english]{babel}
\usepackage{microtype,xspace}
\usepackage[utf8]{inputenc}	

\usepackage{jheppub} 
\usepackage{bm,amssymb,amsmath,amsfonts,slashed,graphicx,multirow,soul,mathtools,xspace,array,comment,color}
\usepackage{hyperref}
\hypersetup{colorlinks=true, allcolors=blue}  
\usepackage{adjustbox}
\usepackage{siunitx}
\usepackage{bm} 
\usepackage{float}
\usepackage{feynmf}
\allowdisplaybreaks
\usepackage{ bbold }
\usepackage{subfigure}
\usepackage[normalem]{ulem} 
\usepackage{enumerate}
\usepackage{multirow}
\usepackage{booktabs}
\usepackage{xparse}
\usepackage{etoolbox}
\usepackage{physics}
\usepackage{hypcap}
\usepackage{siunitx}
\usepackage{tikz}
\usepackage[compat=1.1.0]{tikz-feynman}
\usepackage[capitalise]{cleveref}

\NewDocumentCommand{\lwc}{ m m O{} o }{
	L^{\ifblank{#3}{}{#3,}#2 }_{\IfNoValueTF{#4}{#1}{\substack{#1\\#4}}}
}

\def\be{\begin{equation}}
\def\ee{\end{equation}}
\newcommand{\TeV}{\text{TeV}}
\newcommand{\GeV}{\text{GeV}}
\newcommand{\Zga}{Z/\gamma}
\newcommand{\BFmu}{{\bf \mu}}
\newcommand{\BFmubar}{{\bf \bar \mu}}

\arxivnumber{} 

\title{On the impact of the mixed $Z/ \gamma$ PDF at muon colliders}

\author[1]{David Marzocca}
\author[1,2]{, Alfredo Stanzione}

\affiliation[1]{INFN, Sezione di Trieste, SISSA, Via Bonomea 265, 34136, Trieste, Italy}
\affiliation[2]{SISSA International School for Advanced Studies, Via Bonomea 265, 34136, Trieste, Italy}

\emailAdd{david.marzocca@ts.infn.it}
\emailAdd{alfredo.stanzione@sissa.it}

\abstract{
We study the role of the $Z/\gamma$-interference parton distribution function (PDF) in high-energy muon colliders.
We review how this PDF emerges when electroweak interactions are applied to the collinear splitting process and show that the leading-order approximation is significantly suppressed due to an accidental cancellation. 
However, this suppression does not appear in the leading-logarithm resummed numerical result, where the $Z/\gamma$ PDF is instead comparable to those of other electroweak gauge bosons.
By extending the analytical approximation to next-to-leading order, we show the mechanism by which the suppression is lifted and provide a more accurate approximation to the numerical result.
Furthermore, we explore the impact of the $Z/\gamma$ PDF in several processes at future muon colliders. High-energy Compton scattering is identified as a promising process for observing experimentally this peculiar electroweak effect with high precision. We also quantify the impact of the $Z/\gamma$ PDF on Higgs physics and, as a new physics example, in resonant single-production of axion-like particles (ALP).}

\begin{document}
\maketitle
\flushbottom

\section{Introduction}

The Standard Model (SM) of particle physics has been remarkably successful, providing accurate predictions across a wide range of processes up to the $\mathcal{O}(1\,\TeV)$ energy scale. Beyond this scale, the potential for new physics with $\mathcal{O}(1)$ couplings to SM fields remains an open, well motivated, and exciting possibility. Searching for such beyond-the-SM physics is a central objective of the Large Hadron Collider (LHC) and future high-energy colliders.

However, the TeV energy range is not just significant for the potential discovery of new physics. At these scales, the effects of electroweak (EW) symmetry breaking diminish, leading to an effective restoration of the EW gauge symmetry. This restoration introduces a host of complex phenomena, such as Sudakov double logarithmic corrections \cite{Amati:1980ch,Ciafaloni:1998xg,Ciafaloni:2000gm,Ciafaloni:2000df,Ciafaloni:2000rp,Ciafaloni:2001vt,Ciafaloni:2008cr,Manohar:2018kfx}, EW radiation \cite{Cuomo:2019siu,Chen:2022msz}, EW collinear splitting and EW parton distribution functions (PDFs) \cite{Ciafaloni:2001mu,Ciafaloni:2005fm,Chen:2016wkt,Bauer:2017isx,Bauer:2017bnh,Fornal:2018znf,Cuomo:2019siu,Han:2020uid,Han:2021kes,Azatov:2022itm,Garosi:2023bvq,Ciafaloni:2024alq,Nardi:2024tce}, etc.
While some of these aspects are already relevant at the LHC \cite{Ciafaloni:1998xg,Denner:2000jv,Denner:2001mn,Manohar:2018kfx,Borel:2012by,Pagani:2021vyk,Chay:2022qlc}, future high-energy colliders will probe this energy range directly, making it essential to develop a deep understanding of EW restoration and the related phenomenology to ensure accurate SM predictions.
Among the proposed future colliders, this goal is particularly crucial in the case of TeV-scale muon colliders (MuC) \cite{AlAli:2021let,Stratakis:2022zsk,Jindariani:2022gxj,Aime:2022flm,DeBlas:2022wxr,Accettura:2023ked,Accettura:2024qnk}. In fact, the suppression of QCD effects, due to the non-colored nature of muons, promotes EW interactions to a leading role.

The relevance of EW corrections in multi-TeV muon colliders is exacerbated by the large Sudakov double logarithms, that scale as powers of $\alpha_2 \log^2 E^2 / m_W^2$, where $E$ is the typical energy of the hard process and $\alpha_2$ the weak coupling constant. This factor becomes of $\mathcal{O}(1)$ above the TeV, requiring the resummation of such contributions. These large double logarithms contribute to the process in several ways: to initial-state and final-state radiation (ISR and FSR) emission \cite{Chen:2016wkt,Han:2020uid,Han:2021kes,Garosi:2023bvq}, radiative corrections \cite{Ma:2024ayr}, and the soft radiation region \cite{Manohar:2018kfx}. Different approaches can be used to resum these various classes of corrections.
In the case of ISR, this can be achieved via EW PDFs of the muon, obtained as solutions of the differential DGLAP \cite{Gribov:1972ri,Dokshitzer:1977sg,Altarelli:1977zs} equations of collinear splittings.
An accurate SM prediction of any particular process would however require also the inclusion of all the other corrections listed above.
The developement of such a complete framework for dealing with EW corrections, and the understanding of how to properly include EW PDFs and other resummation techniques when comparing theory predictions with experimental observables are some of the main objectives for future theoretical research aimed at deriving accurate SM predictions for muon collider processes.
In this work we focus only on a specific aspect of EW PDFs, with the goal of establishing its quantitative significance at muon colliders.

In a high-energy collision, the emission of ISR can be factorised from the hard scattering process if the emitted radiation is collinear, meaning its transverse momentum is much smaller than the typical energy of the hard scattering due to a small emission angle\footnote{A formalism that allows also to include the soft region of the splitting processes has recently been developed in Ref.~\cite{Nardi:2024tce}.} \cite{Kunszt:1987tk,Borel:2012by,Buttazzo:2018qqp,Cuomo:2019siu,Costantini:2020stv}.
Under this condition, the cross section for the entire process, inclusive over ISR, can be computed by convoluting the partonic cross section of the hard scattering with the PDFs of the corresponding partons in the initial state.
The PDF, $f_i(x, Q^2)$, describes the probability that the parton $i$ carries a longitudinal momentum fraction $x$ of the initial beam momentum, with $Q$ representing the factorization scale.

The PDF formalism, typically employed in the case of proton collisions, can also be used to describe collinear ISR in lepton colliders.
In the case of proton colliders, QCD interactions dominate the phenomenology. Since this interaction becomes non-perturbative at low scales, PDFs of a proton can only be obtained by fitting experimental data.
In lepton colliders, however, at low energy scales the leading interaction is QED, allowing for a derivation of PDFs of leptons from first principles by solving the DGLAP equations with boundary conditions set at the lepton mass scale \cite{Frixione:2019lga}.
When the factorisation scale $Q$ rises above the EW scale, QED interactions should be substituted with the complete EW ones in order to resum also the possibly large logarithms due to collinear emission of electroweak radiation \cite{Ciafaloni:2005fm}.
EW interactions introduce several new phenomena in the evolution of PDFs. Some of the most relevant ones are: Sudakov double-logs \cite{Ciafaloni:1998xg,Ciafaloni:2000df,Ciafaloni:2001vt}, polarization effects \cite{Bauer:2018arx}, ultra-collinear splittings \cite{Chen:2016wkt}, and EW mass effects \cite{Chen:2016wkt,AlAli:2021let}.
\begin{figure}[t]
\centering
\includegraphics[width=10cm]{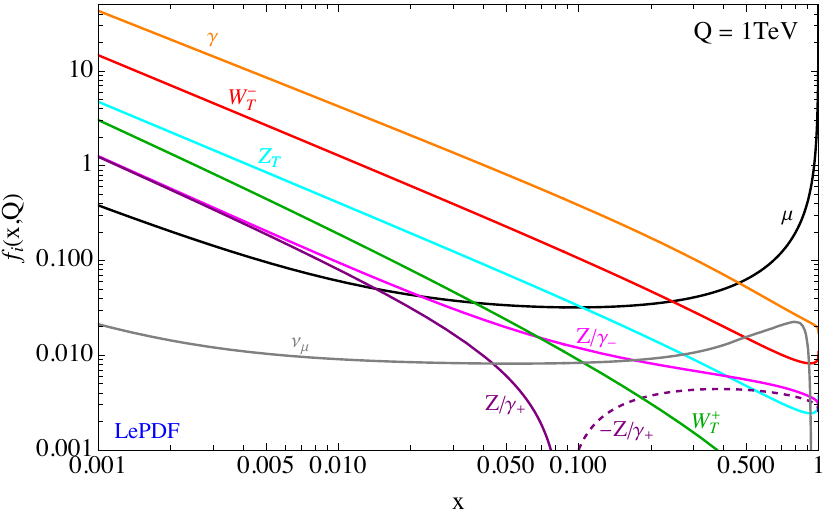}
	\caption{Examples of some SM PDFs of a muon obtained with LePDF \cite{Garosi:2023bvq}, for a factorization scale $Q = 1 \, \TeV$. Except for the $Z/\gamma$ PDF, the other are summed over the transverse helicities. The dashed line indicates that the PDF is negative, in which case we plot the absolute value.}
 \label{fig:LePDF}
\end{figure}
The leading-logarithmic resummation of the collinear splittings leading to SM PDFs can be obtained  by solving the SM DGLAP equations \cite{Ciafaloni:2001mu,Ciafaloni:2005fm,Chen:2016wkt,Bauer:2017isx,Bauer:2017bnh,Cuomo:2019siu,Han:2020uid,Han:2021kes,Azatov:2022itm,Garosi:2023bvq}, a large set of coupled differential equations whose solution encode all the information about initial state radiation with multiple collinear emissions:
\begin{equation}
    Q^2\frac{df_B(x,Q^2)}{d Q^2} = P_B^v \, f_B(x,Q^2)+\sum_{A,C}\frac{\alpha_{ABC}}{2\pi} \widetilde{P}_{BA}^C \otimes f_A + \frac{v^2}{16\pi^2Q^2}\sum_{A,C}\widetilde{U}_{BA}^{C}\otimes f_A ~,
    \label{eq:DGLAP_SM}
\end{equation}
where $P_B^v$ describes the virtual corrections for the parton $B$ and $\alpha_{ABC} \widetilde{P}_{BA}^C$ ($v^2 \widetilde{U}_{BA}^{C}$) describe the (ultra-)collinear splitting process $A \to C B$ in the case of massive partons, which are convoluted with the $f_A$ PDFs. We refer to Ref.~\cite{Garosi:2023bvq} for further details on the formalism and the complete numerical solutions of these equations, which we employ in the following. In \cref{fig:LePDF} we show some examples of SM PDFs of a muon obtained with LePDF.

A peculiar feature of SM PDFs is the necessary presence of PDFs describing mixed-states, due to the possible interference between the photon and the transverse polarizations of the $Z$ boson, $f_{Z/\gamma_{\pm}}(x,Q^2)$, as well as between its longitudinal polarization and the Higgs \cite{Ciafaloni:2000gm,Ciafaloni:2005fm,Chen:2016wkt}.
Such mixed PDFs are not present in QCD or QED, upon integrating over the azimutal angle of the emitted collinear radiation \cite{Cuomo:2019siu}, and is therefore a specific EW effect, due to the fact that those states have the same quantum numbers under the unbroken gauge group.
In our paper we study the impact of the mixed $\Zga$ PDF in MuC phenomenology.
In \cref{sec:ZgaPDF} we review the $\Zga$ PDF and show that the leading-order approximation is not suitable for an accurate description. On the other hand, extending the calculation to $\mathcal{O}(\alpha^2)$ allows us to derive an analytic expression that is in good agreement with the numerical result obtained in Ref.~\cite{Garosi:2023bvq} by resumming at the leading-log order the full set of SM DGLAP equations.
In \cref{sec:Compton} we discuss a process, namely high-energy Compton scattering, that could allow to measure experimentally the impact of the $\Zga$ PDF, and we quantify the potential precision attainable.
Then, in \cref{sec:WHproduction} we study its effect in $WH$ production, while in \cref{sec:BSMsinglet} we show how also some new physics searches can be affected by this mixed PDF by focusing on single resonant production of axion-like scalar singlets (ALP).
Finally, we conclude in \cref{sec:conclusions}.
Details of our computations are collected in the Appendices.

\section{The mixed $Z_T/ \gamma$ PDF}\label{sec:formalism}
\label{sec:ZgaPDF}

\begin{figure}[t]
\centering
\begin{tikzpicture}
  \begin{feynman}
    \vertex (a) at (0,0) {\small \(A\)};
    \vertex (b) at (2,0);
    \vertex (f1) at (4,0.4) {\small \(C\)};
    \vertex[blob,minimum size=1.3cm] (b3) at (3.5,-1.2) {};
    \vertex (d) at (1,-2) {\small \(X\)};
    \vertex (y1) at (5.2,-0.8);
    \vertex (y2) at (5.2,-1.2);
    \vertex (nome) at (5.5,-1.2) {\small \(Y\)};
    \vertex (y3) at (5.2,-1.6);
    
    \diagram* {
      (a) -- (b) -- (f1),
      (b) -- [edge label'=\small\(B\)] (b3),
      (b3) --  (d), 
      (b3) -- (y1),
      (b3) -- (y2),
      (b3) -- (y3),
    };
  \end{feynman}
\end{tikzpicture}
\caption{\label{fig:AX>CY} Schematic diagram for the process $A X \to C Y$, with initial-state collinear splitting $A \to B C$, followed by a hard scattering $B X \to Y$.}
\end{figure}
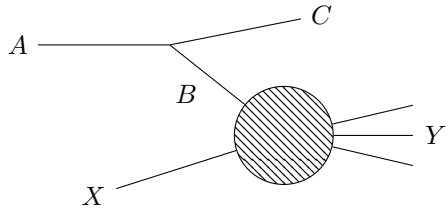

Let us start by reviewing briefly some implications of collinear factorisation, focussing on the initial-state splitting processes and following the discussion of Ref.~\cite{Cuomo:2019siu}.
Consider a process $A X \to C Y$, which can proceed through the exchange of a virtual particle $B^*$ with electroweak-scale mass $m$ as the emission $A \to C B^*$, with $C$ having a transverse momentum $|{\bf k}_\perp|$ relative to the direction of $A$, followed by the hard scattering $B^* X \to Y$ of typical hard energy $E$, as depicted schematically in \cref{fig:AX>CY}.
Collinear factorization states that if $\delta_m = m / E \ll 1$ and $\delta_\perp = |{\bf k}_\perp| / E \ll 1$ then the amplitude factorizes as
\be
    i \mathcal{M}(A X \to C Y) = \sum_B i \mathcal{M}^{\rm split}(A \to C B^*) \frac{i}{Q^2} i \mathcal{M}^{\rm hard}(B X \to Y) \left(1 + \mathcal{O}(\delta_{m,\perp}) \right)~,
    \label{eq:factorization}
\ee
where in the hard scattering matrix element the state $B$ is taken as approximately on-shell.
The differential cross section $d\sigma$ is proportional to the modulus square of \cref{eq:factorization} and therefore it contains the interference terms of different states $B$ and $B^\prime$ that can enter the same splitting and hard processes, which in general could be different species or different helicities:
\be
    d \sigma  
        \propto \left| \mathcal{M}(A X \to C Y) \right|^2 
        \propto \sum_{B, B^\prime} d\rho^{\rm split}_{B B^\prime} \, d\rho^{\rm hard}_{B^\prime B} = \Tr \left[ d\rho^{\rm split} d\rho^{\rm hard} \right]~,
\ee
where the density matrices for the splitting and hard processes are proportional to
\be\begin{split}
    d\rho^{\rm split}_{B B^\prime} &\propto  \Re\left[\mathcal{M}^{\rm split}(A \to C B^*) \mathcal{M}^{\rm split}(A \to C B^{\prime *})^* \right]~, \\
    d\rho^{\rm hard}_{B^\prime B} &\propto \Re\left[\mathcal{M}^{\rm hard}(B^\prime X \to Y)^*  \mathcal{M}^{\rm hard}(B X \to Y) \right]~,
\end{split}\ee
and we refer to \cite{Cuomo:2019siu} for details. Upon integrating over the azimutal angle of the collinear emission process, the interference between different helicity states vanishes.\footnote{As pointed out in \cite{Borel:2012by,Cuomo:2019siu}, upon integrating over the splitting azimutal angle the corrections to the factorization expression become of $\mathcal{O}(\delta^2_{m, \perp})$.}

In order for the interference to be non-vanishing, the states $B$ and $B^\prime$ should be interchangable in both the splitting and hard processes.
This implies they should have the same conserved charges, i.e. same electric charge and color representation, but also same family lepton number $L_{e, \mu, \tau}$, baryon number and, if CKM-suppressed splitting processes are neglected, also individual baryon number for each generation $B_{1,2,3}$. With these constraints, in the SM the only possible non-vanishing interference terms are between the photon and the transverse $Z$ boson, inducing the mixed $Z/\gamma$ PDF, or between the longitudinal $Z$ boson and the phyiscal Higgs, responsible for the mixed $h/Z_L$ PDF \cite{Ciafaloni:2000gm,Ciafaloni:2005fm,Chen:2016wkt,Cuomo:2019siu}.
The mass difference between the two states is at most of order $m$, i.e. the EW scale, therefore the effects due to the different virtuality will be of $\mathcal{O}(\delta_m^2)$, compatible with the approximation on which factorization is based.
In this work we focus on the mixed $Z/\gamma$ case, since its effects are much larger than those due to the $h/Z_L$ splitting.

\subsection{Comparison with the Effective Vector Boson Approximation}\label{sec:NLO}

Approximate analytical solutions to the DGLAP equations can be obtained iteratively by solving them order-by-order in the coupling expansion, starting from the zeroth-order result $f_{\mu_L}^{(0)}(x,Q^2)=f_{\mu_R}^{(0)}(x,Q^2)=\frac{1}{2}\delta (1-x)$ and $f_{i \neq \mu_{L,R}}^{(0)}(x,Q^2) = 0$.
At the first order, these zeroth-order results are substituted in the right-hand-side of \cref{eq:DGLAP_SM}. In this way a single collinear splitting is considered and all the differential equations decouple, allowing for a straightforward analytical solution.
This procedure was applied to the EW gauge bosons PDFs, resulting in what is also known as Effective Vector Boson Approximation (EVA) \cite{Fermi:1924tc,vonWeizsacker:1934nji,Williams:1934ad,Landau:1934zj,Cahn:1983ip,Dawson:1984gx,Chanowitz:1984ne,Kane:1984bb,AlAli:2021let,Ruiz:2021tdt}.
Since only one collinear emission is considered, the approximate solution is of linear $\mathcal{O}(\alpha)$ order in the coupling. 

\begin{figure}[t]
\includegraphics[scale=0.52]{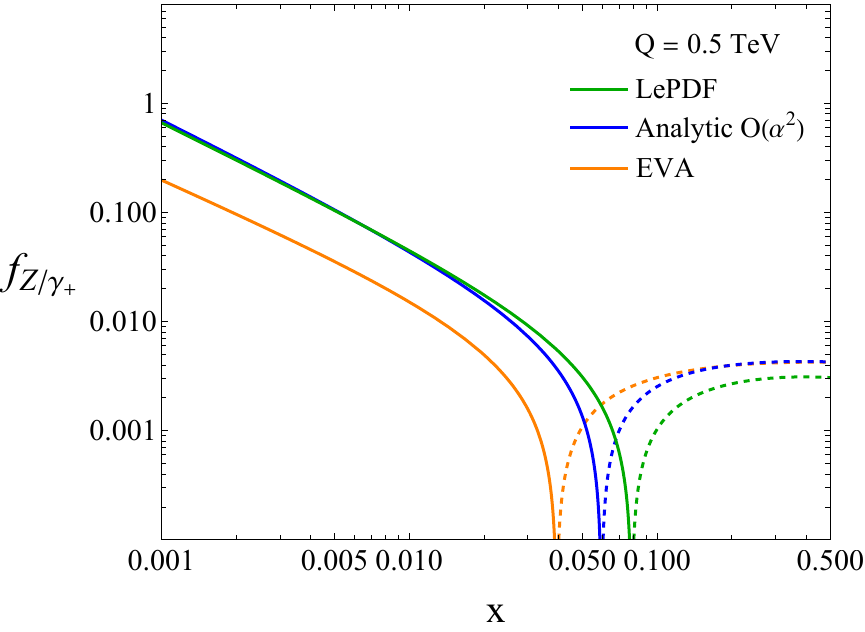}
\includegraphics[scale=0.52]{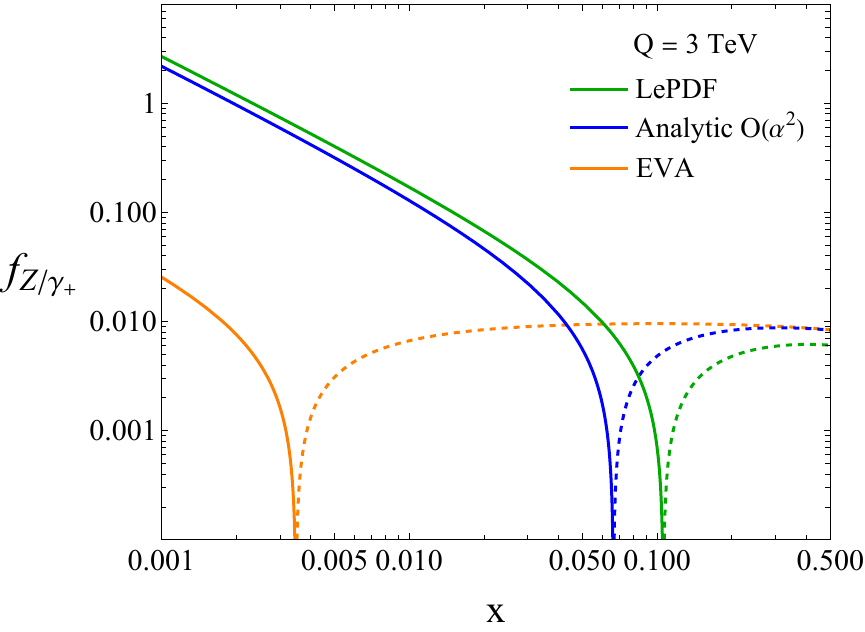}\\
\includegraphics[scale=0.52]{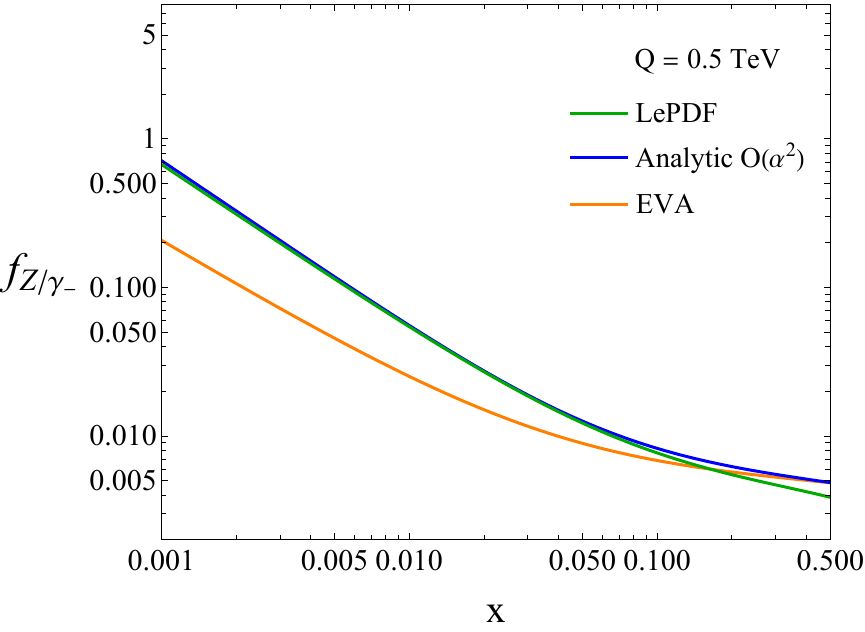}
\includegraphics[scale=0.52]{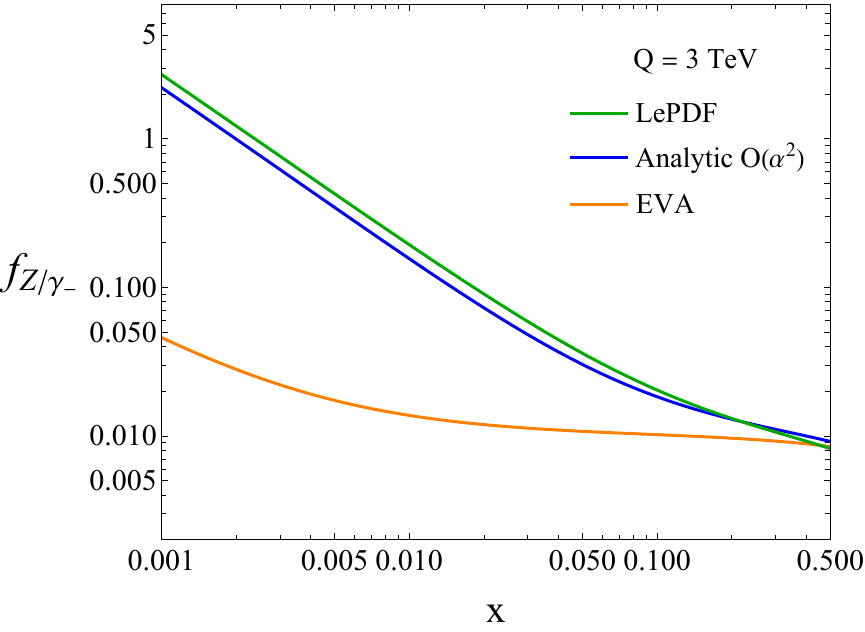}
	\caption{Comparison of the EVA $\mathcal{O}(\alpha)$ approximation in orange, the $\mathcal{O}(\alpha^2)$ approximation in blue and the numerical full solution to the DGLAP equation in green. Dashed lines represent negative values. In the upper (lower) panels, we show the $Z/\gamma$ PDF for $+(-)$ helicity. In both cases, the factorization scale is fixed to $Q=0.5$ TeV (left) and 3 TeV (right).
    }
 \label{fig:comparison}
\end{figure}

Following this procedure, the EVA computation can be carried out for the mixed $Z/\gamma$ contribution as well, obtaining
\be
\setlength{\jot}{10pt}
\begin{split}
f^{(\alpha)}_{Z/\gamma_{\pm}}(x,Q^2)
&= -\int_{m_{\mu}^2}^{Q^2}dp_T^2 \,\frac{\alpha_{\gamma 2}}{2 \pi c_W}\frac{1}{(p_T^2 + (1-x)m_Z^2)} \Bigl( P^{f}_{V_{\pm} f_L}(x) Q_{\mu_L}^Z + P^{f}_{V_{\pm} f_R}(x)Q_{\mu_R}^Z \Bigr)= \\
& =-\frac{\alpha_{\gamma 2}}{2 \pi c_W} \Big( P^{f}_{V_{\pm} f_L}(x) Q_{\mu_L}^Z + P^{f}_{V_{\pm} f_R}(x) Q_{\mu_R}^Z \Big)\, \textrm{log}\frac{Q^2+(1-x)m_Z^2}{m_{\mu}^2+(1-x)m_Z^2}  \,, 
\label{eq:fZgammaEVA}
\end{split}
\ee
where by the $\pm$ subscript we denote the helicity and the splitting functions are 
\be
P^{f}_{V_{+} f_L}(x)=P^{f}_{V_{-} f_R}(x)=\frac{(1-x)^2}{x}\,,\qquad \qquad P^{f}_{V_{-} f_L}(x)=P^{f}_{V_{+} f_R}(x)=\frac{1}{x}\,.
\ee
For brevity, we defined $\alpha_{\gamma 2}=\sqrt{\alpha_{\gamma}\alpha_2}$,  $Q^Z_{\mu_L} = - \frac{1}{2} + s_W^2$, $Q^Z_{\mu_R} = s_W^2$, and $s_W$ is the sine of the Weinberg angle.
As it was also pointed out in Ref.~\cite{Garosi:2023bvq}, this EVA result for the $Z/\gamma$ PDF is accidentally suppressed due to the fact that, for $x\ll 1$, it is proportional to $Q_{\mu_L}^{Z}+Q_{\mu_R}^{Z}=-\frac{1}{2}+2s_W^2 \approx -0.038$,
where we used $s_W^2 (m_Z) \approx 0.231$ (at higher renormalization scales the $\overline{\rm{MS}}$ Weinberg angle grows, making the cancellation even stronger). 
This well known accidental suppression, by at least one order of magnitude, of the vectorial couplings of a charged lepton to the $Z$ boson enters our result because in the initial conditions $f_{\mu_{L,R}}^{(0)}$ it was assumed that the initial muon beam is not polarized. 
Conversely, this tuned cancellation is lifted in the full numerical evolution since a polarization is induced in the muon PDF by EW interactions and, more importantly, contributions from multiple splittings with other particles are considered. Eventually, the full numerical solution exhibits an enhancement of up to two orders of magnitude \cite{Garosi:2023bvq}, as can be noticed by comparing the orange and green lines in \cref{fig:comparison}.

The origin of this enhancement can be understood also analytically, already at $\mathcal{O}(\alpha^2)$. For this purpose we solve iteratively the DGLAP equations at second order, using this time the $\mathcal{O}(\alpha)$ solutions for fermion and gauge boson PDFs appearing in the r.h.s. of \cref{eq:DGLAP_SM}:\footnote{Note that the $W^+$ PDF receives no contributions at order $\alpha$. }
\be\begin{split}\label{eq:a2split}
\frac{df^{(\alpha^2)}_{Z/\gamma_{+}}(x,Q^2)}{dt} \,= & \,  \frac{\alpha_{\gamma2}(t)}{2\pi}2c_W P^{V}_{V_+ V_{\pm}}\otimes f_{W_{\pm}^{-}}^{(\alpha)}+\frac{\alpha_{\gamma 2} (t)}{2 \pi}\frac{c_{2W}(t)}{c_W(t)}P^{h}_{V_+ h}\otimes f_{W^-_L}^{(\alpha)}+ \\
&+\frac{\alpha_{\gamma2}(t)}{2\pi}\frac{2}{c_W(t)}\sum_{f}Q_f\left[ Q^Z_{f_L} P^{f}_{V_+ f_L}\otimes f^{(\alpha)}_{f_L} + Q^{Z}_{f_R} P^{f}_{V_- f_L} \otimes f^{(\alpha)}_{f_R}\right]\,,
\end{split}\ee
where we defined the evolution variable $t=\textrm{log}(Q^2/m_{\mu}^2)$.
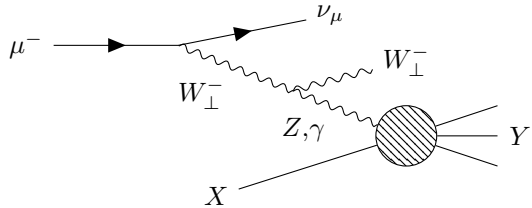
\begin{figure}[t]
\centering
\begin{tikzpicture}
  \begin{feynman}
    \vertex (a) at (0,0) {\small \(\mu^{-}\)};
    \vertex (b) at (2,0);
    \vertex (f1) at (4,0.4) {\small \(\nu_{\mu}\)};
    \vertex (c) at (3.5,-0.6);
    \vertex (b2) at (5,-0.2) {\small \(W^-_{\perp}\)};
    \vertex[blob,minimum size=0.8cm] (b3) at (5,-1.2) {};
    \vertex (d) at (2.5,-2) {\small \(X\)};
    \vertex (y1) at (6.2,-0.8);
    \vertex (y2) at (6.2,-1.2);
    \vertex (nome) at (6.5,-1.2) {\small \(Y\)};
    \vertex (y3) at (6.2,-1.6);
    
    \diagram* {
      (a) -- [fermion] (b) -- [fermion] (f1),
      (b) -- [boson, edge label'=\small\(W^{-}_{\perp}\)] (c),
      (c) -- [boson] (b2),
      (c) -- [boson, edge label'=\small\(Z\textrm{,}\gamma\)]  (b3) --  (d), 
      (b3) -- (y1),
      (b3) -- (y2),
      (b3) -- (y3),
    };
  \end{feynman}
\end{tikzpicture}
\caption{Sketch of initial state radiation from muon beam. The round blob represents the hard cross section, where the $Z,\gamma$ and their interference enter at order $\alpha^2$ (two splittings). $X$ and $Y$ stand for generic initial and final states involved in the hard scattering.}
\label{fig:Wsplit}
\end{figure}
For example, let focus on the first term in the equation above, corresponding to the double emission of \cref{fig:Wsplit}. The transverse $W^-$ boson PDF is, at leading order\footnote{For the sake of simplicity, we consider any EW splitting process starting from the EW scale $Q_{EW} \equiv m_Z$.}
\be
\setlength{\jot}{10pt}
f_{W^{-}_{\pm}}^{(\alpha)}(x,Q^2)\approx\frac{\alpha_2}{8 \pi} P^{f}_{V_{\pm}f_L}(x) \,\textrm{log} \frac{Q^2}{m_Z^2}\,,
\ee
\emph{i.e.}, it is the EVA approximation derived following similar steps to what we did for \cref{eq:fZgammaEVA} and taking for simplicity the limit of small values of $x$ and large $Q^2$. The convolutions involving $f_{W^{-}_{\pm}}^{(\alpha)}$ in \cref{eq:a2split} are computed in detail in \cref{app:NLO}, where the proper treatment of Sudakov double logarithms is discussed. Then, an integral over $t$ is performed starting from the EW scale, that we identify as $m_Z$, up to the factorization scale $Q$. The final analytic result is:
\be\label{eq:fZgammaa2}
\begin{split}
f^{(\alpha^2) P_{VV}}_{Z/\gamma_{+}}(x,Q) = &\,\frac{\alpha_2 \alpha_{\gamma 2}}{96 \pi^2 x}(t-t_Z)^2 \,c_W  \cdot \Big[ 4 (t-t_Z)(1-x)^2 + J(x)\Big]\,,\\[2mm]
f^{(\alpha^2) P_{VV}}_{Z/\gamma_{-}}(x,Q) = &\,\frac{\alpha_2 \alpha_{\gamma 2}}{96 \pi^2 x}(t-t_Z)^2 \,c_W \cdot \Big[ 4(t-t_Z) + K(x)\Big]\,,\\[3mm]
J(x)=&-31 + 60 x - 33 x^2 + 4 x^3 + 12 (1 - x)^2 \log(1 - x) - \\
&-6 (2 - 2 x + x^2) \log(x)\,,\\[2mm]
K(x)=& -31 - 12 x + 39 x^2 + 4 x^3+12 \log(1 - x)-\\
&-6 (2 + 6 x + 3 x^2)\log(x)\,.
\end{split}
\ee
where $t_{Z}=\textrm{log}(m_Z^2/m_{\mu}^2)$. Notice the presence of some double Sudakov logs coming from the $P^{V}_{V_h V_h}\,(h=\pm)$ splittings, namely terms proportional to $\alpha^2 (t-t_Z)^3=\alpha^2 \textrm{log}^3(Q^2/m_Z^2)$, which we find to be the main responsible for the large enhancement in the mixed PDF. The comparison between the EVA estimation of \cref{eq:fZgammaEVA}, the approximate $\mathcal{O}(\alpha^2)$ computation of \cref{eq:a2split}, and the leading-log resummed numerical solution is displayed in \cref{fig:comparison}, where we show the PDFs at fixed $Q=0.5,3$~TeV as function of $x$. The numerical $\mathcal{O}(\alpha^2)$ estimate in the plot contains also the other contributions of \cref{eq:a2split}, whose analytic expressions are fully reported in \cref{app:NLO}.

By comparing the different lines in \cref{fig:comparison} it is clear that higher order splittings do have a large impact on the mixed $f_{Z/\gamma}$, with the main contribution captured by the $W$ boson emission of \cref{fig:Wsplit}, that we computed analytically in \cref{eq:fZgammaa2}. The large difference between the $\mathcal{O}(\alpha)$ and $\mathcal{O}(\alpha^2)$ result could cause concerns regarding possible large corrections at $\mathcal{O}(\alpha^3)$. Such worries can be put to rest by observing that the large $\mathcal{O}(\alpha^2)$ effect is not due to anomalously large corrections but to an anomalously small contribution at leading order (due to the cancellation described above), and by comparing in \cref{fig:comparison} the $\mathcal{O}(\alpha^2)$ line (blue) with the resummed one of LePDF (green).\footnote{For $Z/\gamma_+$, the relative difference between the two increases only near the region where the mixed PDF changes sign, which is to be expected since in that region the PDF is necessarily very small.}

\section{Compton scattering}\label{sec:Compton}

\begin{figure}
\centering

\begin{tikzpicture}
  \begin{feynman}
    \vertex (a1) at (1,0) {\(\mu^{-}\)};
    \vertex (B) at (2.5,-1.2);
    \vertex (B2) at (4,-1.2);
    \vertex (y1) at (5.5,0) {\(\gamma\)};
    \vertex (y2) at (5.5,-2.4) {\(\mu^{-}\)};
    \vertex (b2) at (1,-2.4) {\(Z, \gamma\)};
    
    \diagram* {
      (a1) -- [fermion] (B),
      (B) -- [fermion] (B2),
      (B2) -- [boson] (y1),
      (B2) -- [fermion] (y2),
      (b2) -- [boson] (B),
    };
  \end{feynman}
\end{tikzpicture}
\qquad
\begin{tikzpicture}
  \begin{feynman}
    \vertex (a1) at (1,0) {\(\mu^{-}\)};
    \vertex (B) at (3,-0.8);
    \vertex (B2) at (3,-1.6);
    \vertex (y1) at (5,0) {\(\gamma\)};
    \vertex (y2) at (5,-2.4) {\(\mu^{-}\)};
    \vertex (b2) at (1,-2.4) {\(Z, \gamma\)};
    
    \diagram* {
      (a1) -- [fermion] (B),
      (B) -- [fermion] (B2),
      (B) -- [boson] (y1),
      (B2) -- [fermion] (y2),
      (b2) -- [boson] (B2),
    };
  \end{feynman}
\end{tikzpicture} \\

\begin{tikzpicture}
  \begin{feynman}
    \vertex (a1) at (1,0) {\(\nu_\mu\)};
    \vertex (B) at (2.5,-1.2);
    \vertex (B2) at (4,-1.2);
    \vertex (y1) at (5.5,0) {\(\mu^{-}\)};
    \vertex (y2) at (5.5,-2.4) {\(\gamma\)};
    \vertex (b2) at (1,-2.4) {\( W^- \)};
    
    \diagram* {
      (a1) -- [fermion] (B),
      (B) -- [fermion, edge label'=\small\( \mu^{-} \)] (B2),
      (B2) -- [fermion] (y1),
      (B2) -- [boson] (y2),
      (b2) -- [boson] (B),
    };
  \end{feynman}
\end{tikzpicture}
\qquad
\begin{tikzpicture}
  \begin{feynman}
    \vertex (a1) at (1,0) {\(\nu_\mu\)};
    \vertex (B) at (3,-0.8);
    \vertex (B2) at (3,-1.6);
    \vertex (y1) at (5,0) {\( \mu^{-} \)};
    \vertex (y2) at (5,-2.4) {\(\gamma\)};
    \vertex (b2) at (1,-2.4) {\(W^-\)};
    
    \diagram* {
      (a1) -- [fermion] (B),
      (B) -- [boson] (B2),
      (B) -- [fermion] (y1),
      (B2) -- [boson] (y2),
      (b2) -- [boson] (B2),
    };
  \end{feynman}
\end{tikzpicture}

\caption{Top: leading-order partonic diagrams for Compton scattering at muon \mbox{colliders:} $\mu^- (Z,\gamma) \to \mu^- \gamma$. 
Bottom: leading-order partonic diagrams for the background process $\nu_\mu W^- \to \mu^- \gamma$}
\label{fig:DiagCompton}
\end{figure}
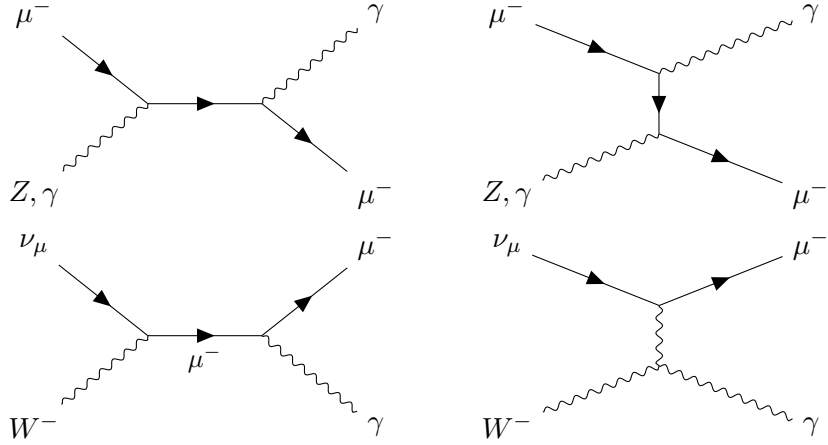

As a first example to showcase the impact of the $Z/\gamma$ PDF we consider the simple SM process of Compton scattering at muon colliders: $\mu^- + (\gamma,Z) \to \mu^- \gamma$, where the initial photon or $Z$ boson comes from the PDF of the anti muon.\footnote{The contribution to the process from picking up the $\mu^-$ PDF inside the anti-muon beam and $\gamma,Z$ from the muon is negligible.} The leading-order partonic diagrams inducing this process are shown in the top row of \cref{fig:DiagCompton}.
The resulting helicity-dependent partonic cross sections for initial-state photon, $Z$ boson, or mixed $Z/\gamma$ state are reported in \cref{app:xsecCompton}.
This process, where a large new physics effect is not expected due to the strong constraints on the coupling between the muon and the photon or the $Z$ boson,\footnote{The same couplings could be tested also, with better sensitivity to heavy new physics, in $\mu^- \mu^+ \to \mu^- \mu^+$ and $\mu^- \mu^+ \to \ell^- \ell^+$, other than in precise low-energy measurements such as that of the muon $(g-2)$.} can be instead a suitable testing ground to study experimentally the effect of the mixed $Z/\gamma$ PDF, by comparing the experimental results with the SM prediction. This is the perspective under which we study it, with the goal of obtaining the size of this novel contribution and estimating the potential experimental reach to measuring it.

\begin{figure}[t]
    \centering
    \includegraphics[height=6.5cm]{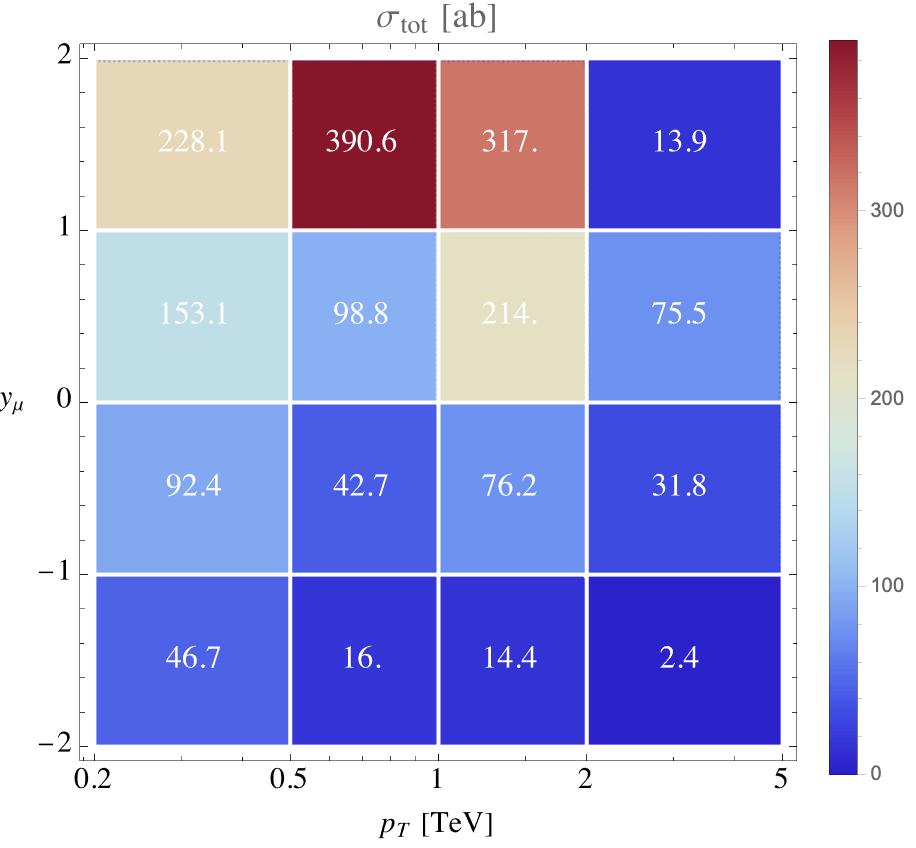}
    \quad
    \includegraphics[height=6.5cm]{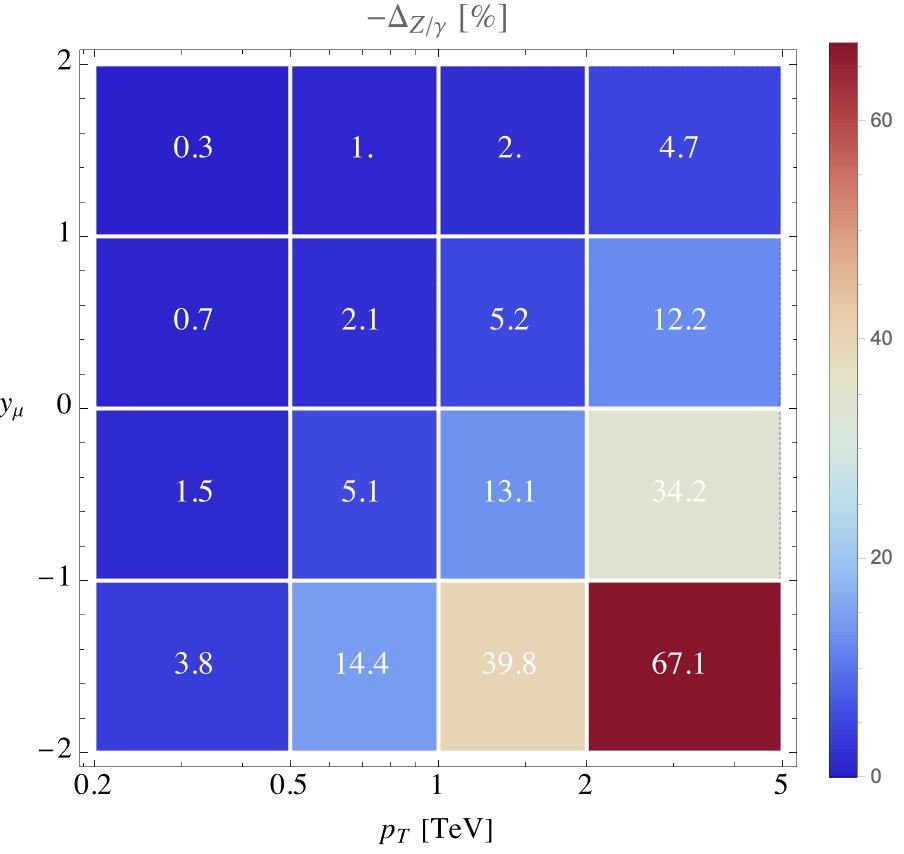}\\
	\caption{In the left panel we show the total cross section, in ab, for $\mu^- \gamma$ production at a 10 TeV MuC in bins of muon $p_T$ and rapidity $y_\mu$, integrating in the photon rapidity between $y_\gamma \in [-2,2]$. The right panel shows the $\Delta_{\Zga}$ ratio in percent for the same bins.}
    \label{fig:compton}
\end{figure}

To obtain cross section for $\mu^- \gamma$ production at a muon collider we multiply the partonic cross sections with the corresponding PDFs, as described in \cref{app:diffxsec}.
To the partonic process of interest, however, we must also add the process $\nu_\mu W^- \to \mu^- \gamma$, since it has the same final state (see bottom row of \cref{fig:DiagCompton}). We expect this to be subleading due to the suppression of the initial state's PDFs, compared to the one of $\mu^-$ and $\gamma$ or $Z$.
The sum of all contributions gives us the fully differential cross section in the muon and photon rapidities and their $p_T$.
In \cref{fig:compton} (left panel) we show the total cross section (in ab) in bins of $p_T$ and muon rapidity $y_\mu$, where we integrated over the photon rapidity between -2 and 2. We recall that the planned integrated luminosity for a 10TeV MuC is approximately 10ab$^{-1}$.

\begin{figure}[t]
    \centering
    \includegraphics[width=9cm]{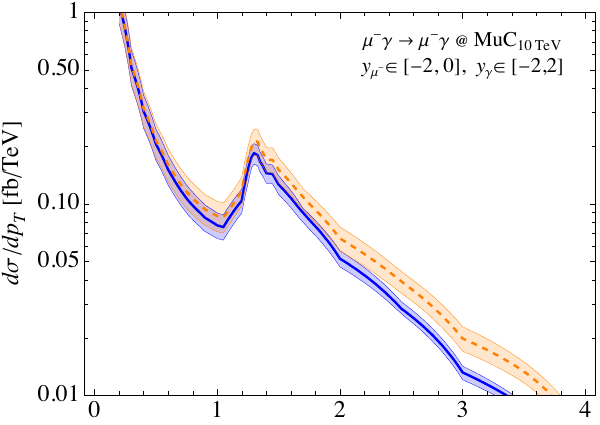}\\[-1mm]
    \hspace{4.5mm}\includegraphics[width=8.7cm]{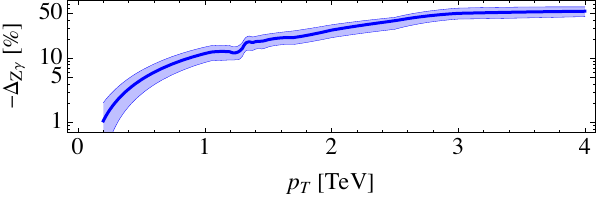}
    \caption{Differential cross section $d\sigma / d p_T$ for $\mu^- \gamma$ production at a 10TeV MuC, after having integrated over $y_\gamma \in [-2,2]$ and $y_\mu \in [-2, 0]$. The solid blue (dashed orange) line include (exclude) the contribution from the $Z/\gamma$ PDF. Its relative contribution, in percentage, is shown in the lower panel. The colored bands are obtained by varying the factorizaton scale in the PDFs as $Q \in [p_T/2, 2 p_T]$.}
    \label{fig:compton_RpT}
\end{figure}

To evaluate the impact of the $\Zga$ PDF we define the ratio $\Delta_{\Zga}$ of the $Z/\gamma$-PDF contribution to the total cross section, over the full result:
\be
    \Delta_{\Zga} \equiv \frac{\sigma_{\Zga}}{\sigma_{\rm tot}}~,
    \label{eq:DeltaZgamma}
\ee
where the ratio is taken either for the same bin or at the level of differential cross sections, depending on the plot.
We observe that this ratio is almost independent on $y_\gamma$, which is the reason for integrating over the photon rapidity in \cref{fig:compton}. In the right panel we report $-\Delta_{\Zga}$ for the same bins. 
The impact of the $\Zga$ PDF is negative and ranges in size from about 1\% up to tens of percent at large $p_T$ and backward muons (negative rapidities). We note that the contribution due to the $\nu_\mu W^-$ initial state is about $2\%$ of the total one in the bins with $1 < y_\mu < 2$, less than $1\%$ for $0 < y_\mu < 2$, and at the per-mille level or smaller for $y_\mu < 0$. In particular, it is completely negligible in all the bins where the $Z/\gamma$ PDF gives the largest contribution.

In \cref{fig:compton_RpT} we plot the differential cross section in $p_T$, integrated in rapidities over the full experimental coverage for the photon, $y_\gamma \in [-2,2]$, and for a backward-going muon, $y_\mu \in [-2, 0]$. We show in solid blue (dashed orange) the values obtained by including (excluding) the contribution due to the $\Zga$ PDF.
The $\Delta_{\Zga}$ ratio is shown in the bottom panel.
We estimate the uncertainty in the PDF evaluation by varying the factorization scale around the central value  $Q = p_T$ from $p_T/2$ to $2 p_T$, and report the corresponding uncertainty bands.
We note that the peak at around $p_T \sim 1350 \, \GeV$, that is noticeable in \cref{fig:compton_RpT}, is due to the fact that, for those values of $p_T$ the kinematical configuration with $x_1 = 1$ ($x_1$ being the Bjorken variable for the incoming muon) enters the range of rapidities included in the integration. For $x_1 \approx 1$ the $\mu^-$ PDF gets the large enhancement due to it being the valence parton, remnant of the Dirac delta that describes the zeroth order PDF of the muon.

\begin{figure}[t]
    \centering
    \includegraphics[height=7.5cm]{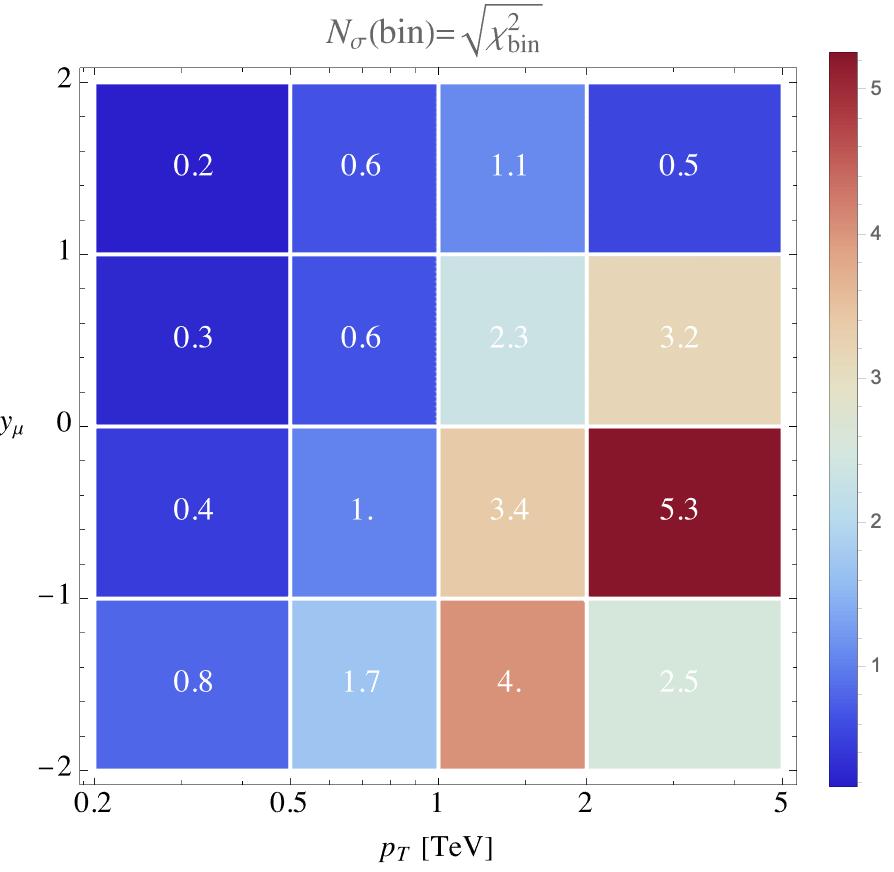}\\
    \caption{Statistical significance of the $\Zga$ contribution to the compton scattering process at a MuC10, in $N_\sigma$, for each $(p_T, y_\mu)$ bin.}
    \label{fig:compton_Nsigma}
\end{figure}

\Cref{fig:compton,fig:compton_RpT} show that this process is very sensitivy to the $\Zga$ PDF, especially going to values of $p_T \gtrsim 500\GeV$ and backward-going muons. 
To better quantify this statement we can calculate the statistical significance of the $\Zga$ contribution over the null assumption, i.e. over the case where it is absent, for each of the $(p_T, y_\mu)$ bins assuming ${\cal L} = 10 \, {\rm ab}^{-1}$ of integrated luminosity by computing\footnote{We only consider statistical uncertainties since they are of several percent in the most sensitive bins: a value larger than the expected future theoretical uncertainties, of about $1\%$.}
\be
    N_\sigma({\rm bin}) \equiv \sqrt{\chi^2_{\rm bin}} \approx \left( {\cal L} \frac{\left( \sigma_{\rm tot} - \hat{\sigma} \right)^2}{\hat{\sigma}} \right)^{1/2}~,
\ee
where $\hat{\sigma} = \sigma_{\rm tot} - \sigma_{\Zga}$~.
The result is shown in \cref{fig:compton_Nsigma}, where we see that in several bins the effect of the $\Zga$ PDF exceeds $3\sigma$ of statistical significance and even $5\sigma$ in the bin $p_T \in [2-5] \TeV$ and $y_\mu \in [-1,0]$.
While at present this effect might be diluted by the PDF scale uncertainty, future theory developements are expected to reduce it substantially since performing precise measurements of EW processes is one of the main goals of a future high-energy muon collider.

\section{Associated Higgs plus $W$ production}
\label{sec:WHproduction}
\label{sec:VBF}

The large rate of emission of collinear photons and EW gauge bosons from high-energy initial-state leptons leads to the well known statement that \emph{high energy muon colliders are vector boson colliders} \cite{AlAli:2021let,Aime:2022flm,Accettura:2023ked}. Indeed, one of the main processes for the production of electroweakly charged final states in TeV-scale $\BFmu \BFmubar$ collision is through vector boson fusion.
When the typical energy of the hard scattering process is much higher than the EW scale then collinear factorization can be applied and the process can be described in terms of the gauge bosons PDFs.
While the most important channel for Higgs production at a MuC is single-production \cite{Forslund:2022xjq,Aime:2022flm,Accettura:2023ked}, associate production with a $Z$ or $W$ boson is also relevant and can offer additional handles to constrain Higgs couplings.

Our goal is to quantify the impact of the $Z/\gamma$ PDF in $W H$ production and compare it with the expected experimental precision.
Contrary to the Compton scattering studied above, in this case one would like to use precise measurement of this process in order to test for small contributions due to new physics. To do that it is crucial to have a complete and precise SM prediction.

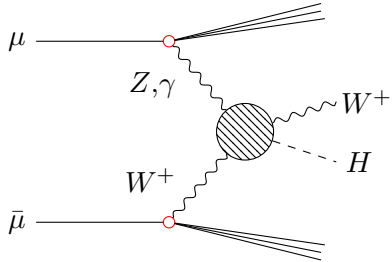
\begin{figure}[t]
\centering
\begin{tikzpicture}
  \begin{feynman}
    \vertex (a1) at (0,0) {\(\mu\)};
    \vertex[empty dot,red,minimum size=0.15cm] (b1) at (2,0) {};
    \vertex (f1) at (4,0.4);
    \vertex (f12) at (4,0.5);
    \vertex (f13) at (4.05,0.3);
    \vertex[blob] (B) at (3,-1.2) {};
    \vertex (y1) at (4.2,-0.8);
    \vertex (W) at (4.6,-0.8) {\(W^{+}\)};
    \vertex (y2) at (4.2,-1.6);
    \vertex (H) at (4.5,-1.6) {\(H\)};
    \vertex (a2) at (0,-2.4) {\(\bar\mu\)};
    \vertex[empty dot,red,minimum size=0.15cm] (b2) at (2,-2.4) {};
    \vertex (f2) at (4,-2.8);
    \vertex (f22) at (4,-2.9);
    \vertex (f23) at (4.05,-2.7);
    
    \diagram* {
      (a1) -- [] (b1) -- [] (f1),
      (b1) -- [] (f12),
      (b1) -- [] (f13),
      (b1) -- [boson, edge label'=\(Z\textrm{,}\gamma\),near start](B),
      (B) -- [boson] (y1),
      (B) -- [scalar] (y2),
      (f2) -- [] (b2) -- [] (a2),
      (f22) -- [] (b2),
      (f23) -- [] (b2),
      (b2) -- [boson, edge label=\(W^{+}\),near start](B),
    };
  \end{feynman}
\end{tikzpicture}
\caption{Sketch of a vector boson fusion process with production of a $W^{+}H$ pair. The internal boson lines stand for $Z,\gamma$ or their interference. The red empty dots emphasize that the boson PDFs are the results to the full DGLAP equations, not only a single splitting.}
\label{fig:WH}
\end{figure}

The cross section for $W^{\pm}H$ production at MuC, sketched in \cref{fig:WH}, has already been investigated, for instance, in Ref.~\cite{Ruiz:2021tdt}. However, these studies employed the EVA approximation for the EW gauge bosons PDFs and totally neglected the  mixed $Z/\gamma$ term.

Total cross sections are computed by integrating the triply differential distribution defined in \cref{app:diffxsec}. For the case under consideration, the parton 2 is a $W^{+}$ while the parton of type 1 can either be either $Z$, $\gamma$, or the interference contribution.
In the latter case, $f_{1}(x_1)=f_{Z/\gamma}(x_1)$ and the hard "cross section" appearing in \cref{eq:TripleDiff} is understood to be proportional to $\Re\left[\mathcal{M}(Z W^{+}\to W^{+} H)\cdot\mathcal{M}(\gamma W^{+} \to W^{+} H)^{*}\right]$, as explained in \cref{sec:formalism}. We compute the differential partonic 
 cross section $d\sigma_{\textrm{H}}/dt$ using the FeynCalc \cite{Shtabovenko:2020gxv,Shtabovenko:2023idz} and FeynArts \cite{Hahn:2000kx} packages. Then, the integrals of \cref{eq:TripleDiff} are evaluated numerically with the following domain restrictions:
\be
|y_W|<2\,,\quad |y_H|<2\,,\quad m>0.5\,\TeV \,,
\ee
where $y$ are the rapidities of final state particles and $m$ their total invariant mass.
The rapidity cuts depend on the detector geometry \cite{Accettura:2024qnk} while the lower value for the energy in the center of mass frame is set in such a way that the factorization assumption is not spoiled.  
In Ref.~\cite{Ruiz:2021tdt}, the authors computed the cross sections with a different choice of cuts, namely $|y_i| < 3, m > 1 \,\TeV$. Employing the same setup, we find a good agreement with their results (see Table 4 in \cite{Ruiz:2021tdt}),  with small deviations ascribable to the use of the complete numerical PDFs.

As this channel could be potentially exploited to look for indirect signatures of new physics, we parameterize possible deviations in the $ZZH$ and $WWH$ couplings and investigate the way they affect the prediction of the total cross section.\footnote{In principle, modified couplings should be employed in the computation of the PDFs as well. However, this effect is negligible since Higgs couplings induce a very small contribution to EW PDFs \cite{Garosi:2023bvq}.}
\begin{figure}[t]
    \centering
    \includegraphics[scale=0.9]{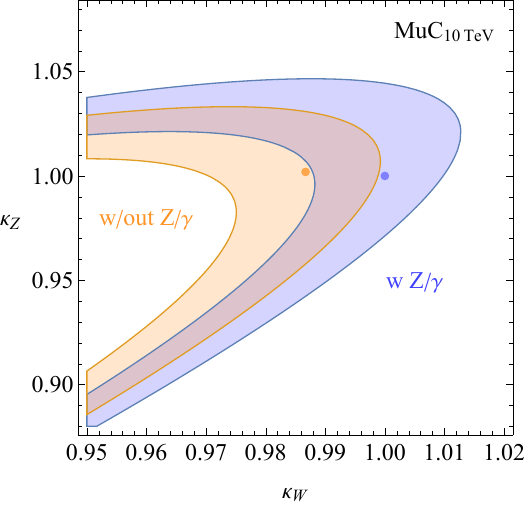}
    \caption{95\% CL bands for the effective couplings $\kappa_{W,Z}$ assuming the SM central value and a relative precision of the 1\% level, for 10 TeV MuC. The factorization scale is $Q=m/2$, where $m$ is the center of mass energy of the hard scattering.}
    \label{fig:fitkWkZ}
\end{figure}
Denoting by $\kappa_{W,Z}$ the multiplicative deviations in the tree-level SM Higgs couplings to EW gauge bosons \cite{LHCHiggsCrossSectionWorkingGroup:2012nn}, such that the SM case is obtained for $\kappa_{W,Z}^{\rm SM} = 1$, and fixing the factorization scale to $Q = m/2$ (we study the impact of scale uncertainties below), we find the following cross sections:
\be
\begin{split}\label{eq:sigmaWH_kappas}
    \sigma_{\textrm{no-} Z/\gamma}^{3\, \TeV} \,\textrm{[fb]}&=12.87 \,\kappa_W^2 + 8.71 \,\kappa_Z^2 -17.79 \,\kappa_W \kappa_Z\,, \\
     \delta\sigma_{Z/ \gamma}^{3\, \TeV} \,\textrm{[fb]} &=-0.075 \,\kappa_W^2-0.010 \,\kappa_W \kappa_Z\,, \\[3mm]
    \sigma_{\textrm{no-} Z/\gamma}^{10\, \TeV} \,\textrm{[fb]}&=135.70 \,\kappa_W^2 + 126.93 \,\kappa_Z^2 -255.82 \,\kappa_W \kappa_Z\,, \\
     \delta\sigma_{Z/ \gamma}^{10\, \TeV} \,\textrm{[fb]} &=-0.15 \,\kappa_W^2-0.030 \,\kappa_W \kappa_Z\,,
\end{split}
\ee
where $\sigma_{\textrm{no-} Z/\gamma}$ is the total cross section computed without taking into account the $Z/\gamma$ PDF. Its contribution is given by $\delta\sigma_{Z/ \gamma}$.
Remarkably, the $Z/\gamma$ interference term mostly modifies the weight of the $\kappa_W$ contribution.
We can have a better insight on its effect by rewriting $\kappa_{W,Z}$ around the SM value, $\kappa_{W,Z} \equiv 1 + \delta_{W,Z}$, and comparing the expressions without and with the mixed PDF:
\be\begin{aligned}
    \sigma_{\textrm{no-} Z/\gamma}^{3\, \TeV} \,\textrm{[fb]}&=
    3.79 + 7.95 \,\delta_W - 0.37 \,\delta_Z 
    + 12.9 \delta_W^2 - 17.8 \delta_W \delta_Z + 8.71 \delta_Z^2 , \\
    \sigma_{\textrm{tot}}^{3\, \TeV} \,\textrm{[fb]}&=
    3.71 + 7.79 \,\delta_W - 0.38 \,\delta_Z 
    + 12.8 \delta_W^2 - 17.8 \delta_W \delta_Z + 8.71 \delta_Z^2 , \\[3mm]
    \sigma_{\textrm{no-} Z/\gamma}^{10\, \TeV} \,\textrm{[fb]}&=
    6.81 + 15.58 \,\delta_W - 1.96 \,\delta_Z 
    + 135.7 \delta_W^2 - 255.8 \delta_W \delta_Z + 126.9 \delta_Z^2 , \\
    \sigma_{\textrm{tot}}^{10\, \TeV} \,\textrm{[fb]}&=
    6.63 + 15.25 \,\delta_W - 1.99 \,\delta_Z 
    + 135.6 \delta_W^2 - 255.9 \delta_W \delta_Z + 126.9 \delta_Z^2 ,
    \label{eq:sigmaWH_deltas}
\end{aligned}\ee
where $\sigma_{\rm tot} = \sigma_{\textrm{no-} Z/\gamma} + \delta\sigma_{Z/ \gamma}$.
From these expressions one can understand that the main effect of the $Z/\gamma$ contribution is to modify the SM prediction, by an amount of approximately 2\% (3\%) for the 3 (10) TeV MuC. Such percent effect is important to be taken into account given the expected future precision in such measurements.
This is illustrated in \cref{fig:fitkWkZ} for the 10 TeV MuC, where we perform a fit in the $(\kappa_W, \kappa_Z)$ plane assuming a measurements of $W H$ production cross section with 1\% precision.\footnote{This is a reasonable estimate given the expected number of events in the $H \to b \bar{b}$ and $W \to \ell \nu$ decay channels, with 10ab$^{-1}$ of integrated luminosity.} For the \emph{truth} value we take the SM point including the $Z/\gamma$ effect, then fit it with the expressions in \cref{eq:sigmaWH_kappas}.
It is worth emphasizing that the $WH$ production receives a background contribution from the muon-neutrino scattering, whose total cross section is $\sigma_{\mu_{L}\,\nu \to W^{+}H}^{3\,\TeV\,(10\,\TeV)}= 0.160\, (0.031)$ fb.\footnote{A dedicated study on the effects due to the muon neutrino PDF inside the muon will be the focus of an upcoming work \cite{Capdevilla:2024zzz}.}

\begin{figure}[t]
    \centering
     \includegraphics[width=10cm]{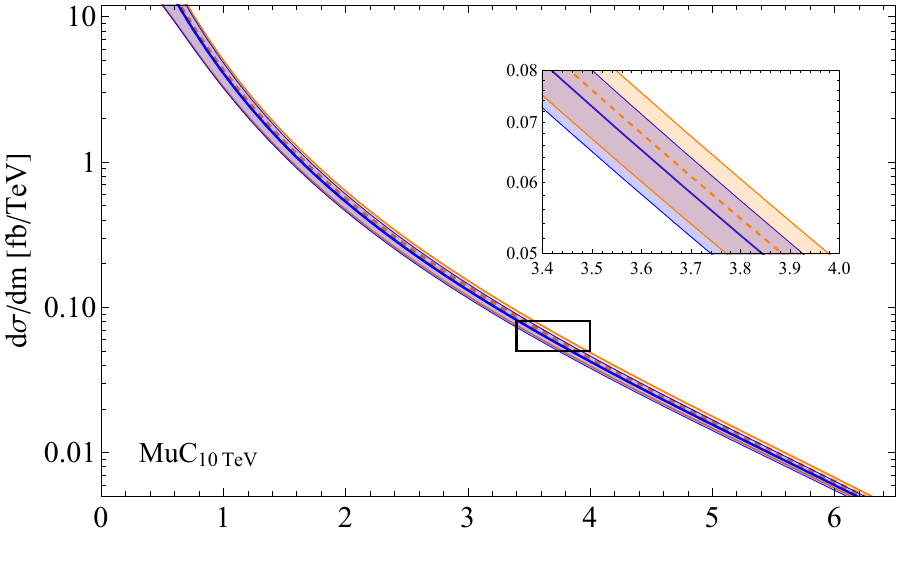}\\[-5.5mm]
    \hspace{2.8mm}\includegraphics[width=9.9cm]{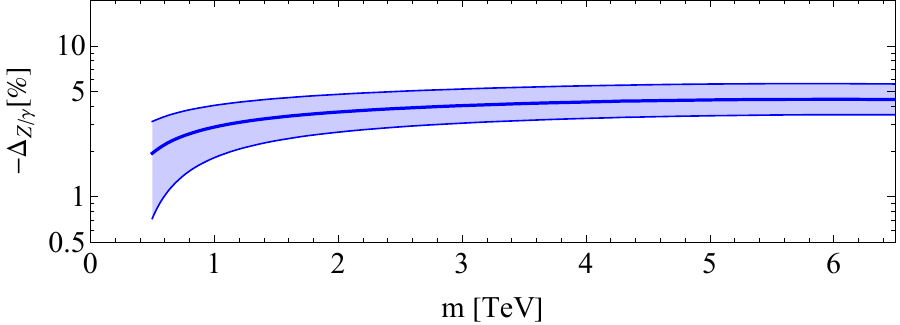}
    \caption{Differential cross section $d\sigma / d m$ for $W^- H$ production at a 10 \TeV\, MuC, after having integrated over $y_{W,H} \in [-2,2]$. The solid blue (dashed orange) line include (exclude) the contribution from the $Z/\gamma$ PDF. Its relative contribution, in percentage, is shown in the lower panel. The colored bands are obtained by varying the factorizaton scale in the PDFs as $Q \in [m/2, 2 m]$.}
    \label{fig:WH_Rm}
\end{figure}

By comparing the cross sections $\sigma_{\textrm{no-} Z/\gamma}$ in \cref{eq:sigmaWH_kappas} with the ones in \cref{eq:sigmaWH_deltas} one notices that near the SM point, $\kappa_{W,Z} \approx 1$, each of the three terms in $\sigma_{\textrm{no-} Z/\gamma}$ is larger than the overall result, signalling that a cancellation takes place in the SM limit. This cancellation is stronger at the 10 TeV MuC than at 3 TeV. Its origin lies in the Higgs's role in unitarising the scattering of longitudinally polarised gauge bosons. In fact, the leading contribution to each of the three terms in $\sigma_{\textrm{no-} Z/\gamma}$ in \cref{eq:sigmaWH_kappas} is due to the scattering of longitudinal helicities: $Z_L W_L^+ \to W_L^+ H$. Away from the SM limit, specifically for values that violate custodial symmetry (i.e. if $\kappa_Z \neq \kappa_W$) the corresponding scattering amplitude grows with the energy and at some point it would violate perturbative unitarity. In the SM limit this energy-growing behaviour is cancelled.\footnote{In the SM Effective Field Theory, a dimension-6 operator that would induce such an effect in this channel is $\mathcal{O}_T = (\Phi^\dagger \stackrel{\leftrightarrow}{D}_\mu \Phi)^2$, where $\Phi$ is the Higgs doublet.}

In \cref{fig:WH_Rm} we plot the SM differential cross section $d\sigma/dm$ for a 10 TeV MuC, as a function of the hard scattering center of mass energy $m$. The colored bands are obtained by varying the factorization scale, akin to what we did in \cref{fig:compton_RpT}. 
Compared with the Compton scattering, the $Z/\gamma$ interference contribution has a weaker effect, so that the colored bands almost superpose. 
The relative size of the effect can be seen in the bottom panel, its size being of a few percent, confirming the result derived above for the integrated cross section.

\section{ALP single production}\label{sec:BSMsinglet}

As an example to showcase the importance of the $Z/\gamma$ PDF also for BSM searches, we consider the process of single production of an axion-like pseudo-scalar particle (ALP), $\phi \sim (1,1)_0$ under the SM gauge group. Neglecting the coupling to gluons, which have a suppressed PDF at muon colliders \cite{Han:2021kes,Garosi:2023bvq}, the leading interactions with electroweak vector bosons are described by the dimension-five operators
\begin{equation}
    \mathcal{L}_{\phi VV}=\frac{C_W}{\Lambda}\phi W_{\mu\nu}^a\widetilde{W}^{\mu\nu,a}+\frac{C_B}{\Lambda}\phi B_{\mu\nu}\widetilde{B}^{\mu\nu}\,.
\end{equation}
Below the electroweak scale, the resulting couplings with physical EW bosons are:
\begin{equation}
    \mathcal{L}_{\phi VV}^{eff}=
    \frac{c_{\phi ZZ}}{4 \Lambda}\phi Z_{\mu\nu}\widetilde{Z}^{\mu\nu}+
    \frac{c_{\phi \gamma \gamma}}{4 \Lambda}\phi F_{\mu\nu}\widetilde{F}^{\mu\nu}+
    \frac{c_{\phi \gamma Z}}{2\Lambda} \phi F_{\mu\nu}\widetilde{Z}^{\mu\nu}+
    \frac{c_{\phi WW}}{2 \Lambda}\phi W_{\mu\nu}^+\widetilde{W}^{-\mu\nu}\,,
\end{equation}
where
\begin{equation}
\begin{split}
    &c_{\phi \gamma \gamma}=4(s_{\theta_W}^2C_{W}+c_{\theta_W}^2C_B),\quad c_{\phi Z Z}=4(c_{\theta_W}^2C_{W}
    +s_{\theta_W}^2C_B),\\
    &\quad c_{\phi \gamma Z}=4 s_{\theta_W} c_{\theta_W}(C_W-C_B),\quad c_{\phi WW}=4 C_W\, .
\end{split}
\end{equation}

Within this model, it is straightforward to compute the partonic cross section for resonant production of a single heavy electroweak singlet in the high energy limit (\emph{i.e.} massless vectors limit) and in the narrow-width approximation ($\Gamma_\phi \ll M_\phi$):
\begin{equation}
    \sigma_H(V_1 V_2 \to \phi)(\hat{s})=\frac{\pi}{4}\frac{c_{\phi V_1 V_2}^2}{\Lambda^2}M_{\phi}^2\delta({\hat{s}-M_{\phi}^2})\,,
\end{equation}
where $V_{1,2} = Z, \gamma$ or $V_{1,2} = W^{\pm}$, and $\hat{s}$ is the energy in the center of mass frame. By angular momentum conservation, only amplitudes with both transverse vectors of same helicity are non-vanishing.
The total cross section for a VBF process at MuC can then be computed by convoluting this with parton luminosities.
\begin{figure}[t]
\centering
\begin{tikzpicture}
  \begin{feynman}
    \vertex (a1) at (0,0) {\(\mu\)};
    \vertex[empty dot,red,minimum size=0.15cm] (b1) at (2,0) {};
    \vertex (f1) at (4,0.4);
    \vertex (f12) at (4,0.5);
    \vertex (f13) at (4.05,0.3);
    \vertex (B) at (3,-1.2);
    \vertex (phi) at (4.3,-1.2) {\(\phi\)};
    \vertex (a2) at (0,-2.4) {\(\bar\mu\)};
    \vertex[empty dot,red,minimum size=0.15cm] (b2) at (2,-2.4) {};
    \vertex (f2) at (4,-2.8);
    \vertex (f22) at (4,-2.9);
    \vertex (f23) at (4.05,-2.7);
    
    \diagram* {
      (a1) -- [] (b1) -- [] (f1),
      (b1) -- [boson, edge label'=\(W\textrm{, }Z\textrm{, }\gamma\),near start](B),
      (b1) -- [] (f12),
      (b1) -- [] (f13),
      (f2) -- [] (b2) -- [] (a2),
      (f22) -- [] (b2),
      (f23) -- [] (b2),
      (b2) -- [boson, edge label=\(W\textrm{, }Z\textrm{, }\gamma\),near start](B),
      (B) -- [scalar] (phi)
    };
  \end{feynman}
\end{tikzpicture}
\caption{Sketch of the vector boson fusion production of a $\phi$ singlet. Interference effects are now possible for both the bosons entering the hard cross section. The red empty dots emphasize that the boson PDFs are the results to the full DGLAP equations, not only a single collinear emission.}
\label{fig:VVphi}
\end{figure}
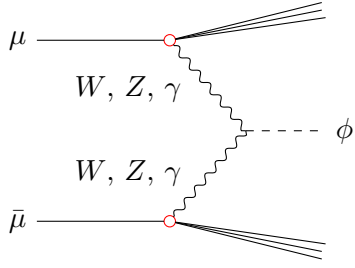
The case of $W^\pm$ in the initial state follows as standard, however for $Z$ and $\gamma$ we should take into consideration also the possible interference in both initial legs, see \cref{fig:VVphi}. To see how this should be implemented it is useful to extend the factorized amplitude of \cref{eq:factorization} to both initial legs:
\be
    i \mathcal{M}(A D \to C E \phi) \propto \sum_{V_1} \sum_{V_2} \frac{\mathcal{M}^{\rm split}(A \to C V_1^*)}{q_1^2 - m_{V_1}^2} \frac{\mathcal{M}^{\rm split}(D \to E V_2^*)}{q_2^2 - m_{V_2}^2} \mathcal{M}^{\rm hard}(V_1 V_2 \to \phi) ~.
    \label{eq:factorization2legs}
\ee
The cross section is obtained by squaring this expression. To describe it, we can introduce two splitting matrices, one each for $\mu$ and $\bar{\mu}$, and define 
\be
\rho_{\mu, \, V_1 V_1^\prime}^{\textrm{split}}= \begin{pmatrix}
f_{\gamma}^{(\mu)} & f_{Z/\gamma}^{(\mu)} \\
f_{Z/\gamma}^{(\mu)} & f_{Z}^{(\mu)} 
\end{pmatrix}\,, \quad 
\rho_{\bar{\mu}, \, V_2 V_2^\prime}^{\textrm{split}}= 
\begin{pmatrix}
f_{\gamma}^{(\bar{\mu})} & f_{Z/\gamma}^{(\bar{\mu})} \\
f_{Z/\gamma}^{(\bar{\mu})} & f_{Z}^{(\bar{\mu})} 
\end{pmatrix}\,,\quad
\sigma^{\textrm{hard}}_{V_1,V_1^{\prime},V_2,V_2^{\prime}} \propto \mathcal{M}^{\textrm{hard}}_{V_1,V_2}\cdot\mathcal{M}^{\textrm{hard} \, *}_{V_1^{\prime},V_2^{\prime}}\,,
\ee
where, given the definition of the couplings and for $M_\phi \gg m_Z$, the hard cross section can be written in full generality as:
\be
\sigma^{\textrm{hard}}_{V_1,V_1^{\prime},V_2,V_2^{\prime}}(\hat{s}) = c_{\phi V_1 V_2} ~ c_{\phi V_1^\prime V_2^\prime} ~ \frac{\pi}{4}\frac{1}{\Lambda^2}M_{\phi}^2 ~ \delta({\hat{s}-M_{\phi}^2})\,.
\ee
The final cross section for this process at a muon collider at energy $\sqrt{s_0}$ is then obtained by summing over the $V_{1,2}^{(\prime)}$ indices, obtaining the right factorization between hard scatterings and PDFs:
\begin{equation}
    \sigma_{\textrm{tot}}^{Z,\gamma}(s_0)= \sum_{V_1, V_1^\prime,V_2, V_2^\prime} \int_{\hat{s}_{min}}^{s_0}d\hat{s} \frac{1}{s_0} \, \mathcal{L}_{V_1,V_1^{\prime},V_2,V_2^{\prime}}(\hat{s},s_0) \cdot \sigma^{\textrm{hard}}_{V_1,V_1^{\prime},V_2,V_2^{\prime}}(\hat{s})~,
\end{equation}
where the parton luminosities are given by
\be
    \mathcal{L}_{V_1,V_1^{\prime},V_2,V_2^{\prime}}(\hat{s},s_0) = \int_0^1 \frac{dz}{z} \rho_{\mu,V_1,V_1^{\prime}}^{\textrm{split}}\left(z, \frac{\hat{s}}{4}\right) \cdot \rho^{\textrm{split}}_{\bar{\mu}, V_2,V_2^{\prime}} \left(\frac{\hat{s}}{z s_0}, \frac{\hat{s}}{4}\right).
\ee
Since in the narrow width approximation the hard cross section is proportional to a Dirac delta, the total cross section simplifies to
\begin{equation}
    \sigma_{\textrm{tot}}^{Z,\gamma}(s_0)= \sum_{V_1, V_1^\prime,V_2, V_2^\prime} \mathcal{L}_{V_1,V_1^{\prime},V_2,V_2^{\prime}}(M_{\phi}^2,s_0) ~ c_{\phi V_1 V_2} ~ c_{\phi V_1^\prime V_2^\prime} ~ \frac{1}{s_0}\frac{\pi}{4}\frac{M_{\phi}^2}{\Lambda^2} ~.
\end{equation}
To obtain the full cross section for single-ALP production, $\sigma_{\rm tot}$, we should then also add the contribution from $W^\pm$ in the initial state, $\sigma_{\rm tot} = \sigma_{\textrm{tot}}^{Z,\gamma} + \sigma_{\textrm{tot}}^{W}$, where
\begin{equation}
    \sigma_{\textrm{tot}}^{W}(s_0)=  \mathcal{L}_{W^- W^+}(M_{\phi}^2,s_0) ~ c_{\phi WW}^2 ~ \frac{1}{s_0}\frac{\pi}{4}\frac{M_{\phi}^2}{\Lambda^2} ~.
\end{equation}

\begin{figure}[t]
	\centering
    \includegraphics[scale=0.51]{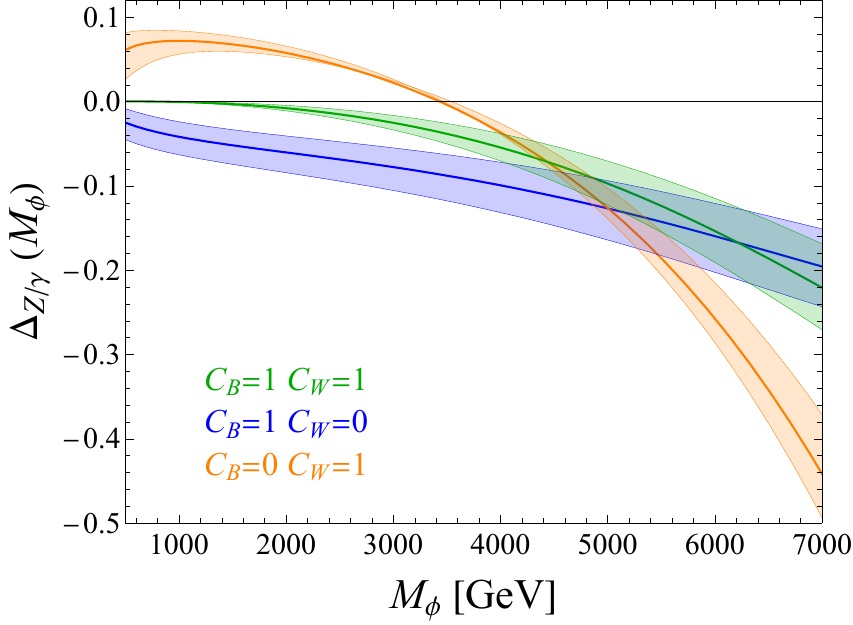}
        \quad
    \includegraphics[scale=0.51]{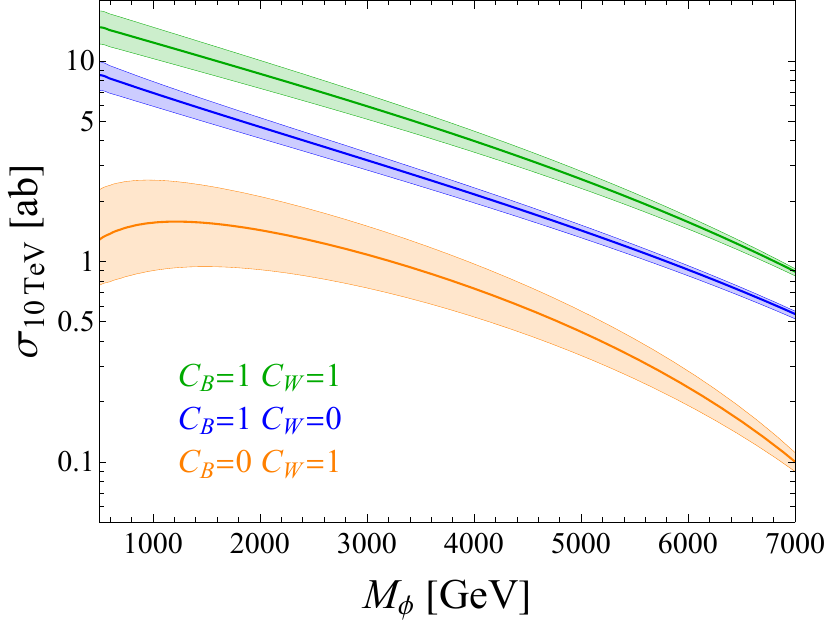}
	\caption{ Left panel: ratio between the interference contribution to the cross section and the total $\sigma_{\textrm{tot}}$. Right panel: total cross section as a function of the singlet mass $M_{\phi}$. Three cases are shown depending on the WCs which are turned on in the UV, where we fix $\Lambda = 1 \TeV$.
 Uncertainty bands correspond to different choices of the factorization scale $ Q=\xi \frac{\hat{s}}{2}$ with $\xi = 0.5, 1, 2$. }
	\label{fig:RphiTOTxsec}
\end{figure}

With the aim of studying the importance of the mixed $\Zga$ PDF, we show in the left panel of \cref{fig:RphiTOTxsec} the ratio between the contribution to $\sigma_{\textrm{tot}}$ due to interference (\emph{i.e.} that would vanish in the $f_{Z/\gamma}\to0$ limit), and the complete $\sigma_{\textrm{tot}}$. Three cases are studied, depending on the interactions that we switch on in the UV. As it is clear from the plot, the effect becomes more and more important at higher mass scales where, however, we expect the total cross section to be suppressed. Indeed, we illustrate in the right panel of \cref{fig:RphiTOTxsec} the total cross section in ab as a function of the EW singlet mass, for $\Lambda =1\,\TeV$. To estimate the mass reach, a reader should keep in mind that at a 10 TeV MuC the integrated luminosity is expected to be 10 ab$^{-1}$.
We observe that, for ALP masses below $\sim 4$ TeV, the $\Zga$ PDF effect ranges from about $+ 10\%$ to $- 10\%$, and gives therefore a sizeable impact in the total production cross section.

\section{Conclusions}
\label{sec:conclusions}

The next generation of multi-TeV colliders offers a unique opportunity to probe EW interactions in a regime where SM gauge symmetries are effectively restored.
Muon colliders, with their high luminosity, large center-of-mass energy available for hard scattering processes, suppressed QCD backgrounds, and relatively small footprint, are ideal for studying these phenomena. 

In this paper, we have provided a detailed analysis of the mixed $Z/\gamma$ PDF, an exotic effect arising from EW interactions in collinear emission of initial-state radiation. 
Our study highlights the suppression of the LO EVA result, due to the well known accidental tuning in the vector-like muon coupling to the $Z$ boson.
By extending the EVA computation to $\mathcal{O}(\alpha^2)$, we have analytically shown how this suppression is lifted, offering a more accurate approximation to the numerical results.

We also explored the phenomenological implications of this PDF at muon colliders. Notably, we identified Compton scattering as a viable process to experimentally detect the $Z/\gamma$ PDF with high precision, potentially achieving a significance greater than $5\sigma$ at a 10~TeV muon collider. Furthermore, we demonstrated that the $Z/\gamma$ PDF can lead to a 2-3\% modification in the SM $W H$ production cross section, which is a substantial effect given the anticipated precision in this channel. Additionally, in the context of new physics, we showed that this PDF could modify the resonant single-production cross section of an ALP by up to 10\% at a 10 TeV muon collider.

The work presented here is a step forward in understanding EW effects in parton distribution functions, a critical component of the broader effort to explore and understand electroweak interactions and their phenomenology in the unbroken regime.
As muon colliders are expected to push the frontiers of high-energy physics, the precise study of such effects will be crucial in refining our understanding of the SM and unlocking new physics.

\section*{Acknowledgments}

The authors acknowledges support by the Italian Ministry of University and Research (MUR) via the PRIN project n.~20224JR28W.

\appendix

\section{Derivation of the $\Zga$ PDF at order $\alpha^2$}
\label{app:NLO}

In this appendix we review the analytical computation of the $f_{Z/ \gamma}(x,Q^2)$ PDF by solving iteratively the DGLAP equations at $\mathcal{O}(\alpha^2)$. We focus here on the positive polarization, but analogous calculation steps can be performed for the negative helicity state, for which the results are reported at the end of the appendix.

We compute the PDF at next-to-leading-order iteratively: we substitute in the DGLAP equation of $f_{Z/ \gamma_+}(x,Q^2)$ the leading order $\mathcal{O}(\alpha)$ expressions of the other PDFs appearing therein. Namely, we study the equation
\be\begin{split}\label{eq:appa2split}
\frac{df^{(\alpha^2)}_{Z/\gamma_{+}}(x,Q^2)}{dt} &\,=  \,  \frac{\alpha_{\gamma2}(t)}{2\pi}2c_W P^{V}_{V_+ V_{\pm}}\otimes f_{W_{\pm}^{-}}^{(\alpha)}+\frac{\alpha_{\gamma 2} (t)}{2 \pi}\frac{c_{2W}(t)}{c_W(t)}P^{h}_{V_+ h}\otimes f_{W^-_L}^{(\alpha)}+ \\
&\,+\frac{\alpha_{\gamma2}(t)}{2\pi}\frac{2}{c_W(t)}(-1)\left[ Q^Z_{\mu_L} P^{f}_{V_+ f_L}\otimes f^{(\alpha)}_{\mu_L} + Q^{Z}_{\mu_R} P^{f}_{V_- f_L} \otimes f^{(\alpha)}_{\mu_R}\right]\,,
\end{split}\ee
where we consider the leading order expressions:
\begin{equation}\label{eq:appEVA}
    \setlength{\jot}{6pt}
    \begin{split}
        & f_{W^{-}_{+}}^{(\alpha)}(x,Q^2)\approx\frac{\alpha_2}{8 \pi} \frac{(1-x)^2}{x} \,(t-t_Z)\,,\\
        & f_{W^{-}_{-}}^{(\alpha)}(x,Q^2)\approx\frac{\alpha_2}{8 \pi} \frac{1}{x} \,(t-t_Z)\,,\\
        & f_{W^{-}_{L}}^{(\alpha)}(x,Q^2)\approx\frac{\alpha_2}{8 \pi} \frac{1-x}{x} \,,\\
        & f_{\mu_L}^{(\alpha)}(x,Q^2)=+\frac{1}{2}\delta(1-x) + \frac{\alpha_2}{16 \pi}\delta(1-x)\left(3 (t-t_Z)-(t-t_Z)^2\right)+\\
        &\qquad\qquad + \frac{1}{4\pi}\left(\frac{1+x^2}{1-x}+\frac{3}{2}\delta(1-x)\right)\left(\alpha_{\gamma}t + \alpha_2\Big(\frac{(1-2 s_W^2)^2}{4 c_W^2}\Big)(t-t_Z)\right)\,,\\
        & f_{\mu_R}^{(\alpha)}(x,Q^2)=+\frac{1}{2}\delta(1-x)+ \frac{1}{4\pi}\left(\frac{1+x^2}{1-x}+\frac{3}{2}\delta(1-x)\right)\left(\alpha_{\gamma}t + \alpha_2\frac{s_W^4}{ c_W^2}(t-t_Z)\right)\,,
    \end{split}
\end{equation}
and we defined $t=\textrm{log}(Q^2/m_{\mu}^2)$ and $t_Z=\textrm{log}(m_Z^2/m_{\mu}^2)$. These leading order formulas can be easily derived and are often used in literature, see e.g. Ref.~\cite{Garosi:2023bvq} for a detailed introduction. 

Regarding the vector boson PDFs, we employ some simplifying approximations. First, all the electroweak ISR effects are considered starting from a unique EW scale, that we identify as $m_Z$ (for both collinear emissions involving $W$ and $Z$ boson vertices). In addition, the above expressions are valid in the limit of high energy $Q^2\gg m_Z^2$ and small $x$. These approximations do not jeopardise the goal of \cref{sec:NLO}, since the most important effects that arise at NLO are anyway captured (see \cref{fig:comparison}), allowing for an analytic understanding of the NLO enhancement. In the rest of the paper, full numerical solutions of DGLAP equations from LePDF \cite{Garosi:2023bvq} were used to compute predictions.

In the next subsections, we discuss in more detail the convolutions appearing in \cref{eq:appa2split}, the origin of the Sudakov logarithms, and we display the full analytic results. We split the computation in three pieces, depending on the states involved in the initial splitting: $P_{VV}^V,P_{Vf}^f,P_{Vh}^h$. We follow the notation and conventions of Ref.~\cite{Garosi:2023bvq}, to which we refer the reader for a complete introduction.

\subsection*{$P_{VV}$ contribution at $\mathcal{O}(\alpha^2)$}

Here we discuss the contribution to \cref{eq:appa2split} due to the collinear emission of a $W_{\pm}^-$ gauge boson, as depicted in \cref{fig:Wsplit}. The convolution with the W boson PDF reads:
\be\label{eq:convPVV}
P^{V}_{V_+ V_{\pm}}\otimes f_{W_{\pm}^{-}}^{(\alpha)}=\int_x^1 \frac{dz}{z}\,\frac{(1-z)^3}{z}\,f_{W_{-}^{-}}^{(\alpha)}\left(\frac{x}{z},Q^2\right)+\int_x^{z_{max}(Q)} \frac{dz}{z}\,\frac{1+z^4}{z(1-z)}\,f_{W_{+}^{-}}^{(\alpha)}\left(\frac{x}{z},Q^2\right)\,.
\ee
Noticeably, following Ref.~\cite{Bauer:2017isx}, we set the upper limit in the second integral to $z_{max}(Q)=1-m_{\textrm{Z}}/Q$. This plays the role of an explicit IR cutoff to the $1/(1-z)$ pole, alternative to the usual $+$-distribution prescription. This procedure can be used in processes where the emitted radiation (e.g. a $W^\pm$ boson) changes the $\text{SU}(2)_L$ component of the initial state and allows to reproduce, to a double log approximation, the Sudakov factor for ISR expected in these cases. More discussions on these effects can be found in \cite{Ciafaloni:2001mu,Bauer:2017isx,Bauer:2017bnh,Amati:1980ch,Manohar:2018kfx}.
The $(1-z)$ pole in this splitting, responsible for the Sudakov log, hints at the importance of the soft region $z\approx 1$. Thus, an interesting future development could be the application of the formalism of Ref.~\cite{Nardi:2024tce} to this calculation.

Upon performing the integrals in \cref{eq:convPVV} and further integrating over $t$, the total contribution to the NLO $Z/\gamma_+$ PDF is:
\be
\begin{split}
f^{(\alpha^2) P_{VV}}_{Z/\gamma_{+}}(x,Q) = &\,
\frac{\alpha_2 \alpha_{\gamma 2}}{96 \pi^2 x}(t-t_Z)^2 \,c_W  \cdot \Big[ 4 (t-t_Z)(1-x)^2 + J(x)\Big]\,,\\[2mm]
J(x)=&-31 + 60 x - 33 x^2 + 4 x^3 + 12 (1 - x)^2 \log(1 - x) - \\
&-6 (2 - 2 x + x^2) \log(x)\,.
\end{split}
\ee
The $\alpha^2 (t-t_Z)^3$ terms appearing in this formula contain the Sudakov double logs mentioned above. Due to the gap between the EW scale and the multi-TeV energy scales explored at a MuC, these additional powers of $\textrm{log}(Q^2/m_{Z}^2)$ provide a remarkable enhancement. The contribution described in this section is indeed the largest one appearing at $\alpha^2$ order.

\subsection*{$P_{Vf}$ contribution at $\mathcal{O}(\alpha^2)$}
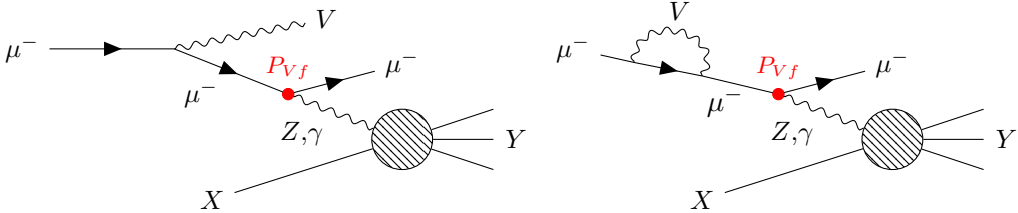
\begin{figure}[th]
\centering
\begin{tikzpicture}
  \begin{feynman}
    \vertex (a) at (0,0) {\small \(\mu^{-}\)};
    \vertex (b) at (2,0);
    \vertex (f1) at (4,0.4) {\small \(V\)};
    \vertex[dot,red,minimum size=0.15cm] (c) at (3.5,-0.6) {};
    \vertex (b2) at (5,-0.2) {\small \(\mu^{-}\)};
    \vertex[blob,minimum size=0.8cm] (b3) at (5,-1.2) {};
    \vertex (d) at (2.5,-2) {\small \(X\)};
    \vertex (y1) at (6.2,-0.8);
    \vertex (y2) at (6.2,-1.2);
    \vertex (nome) at (6.5,-1.2) {\small \(Y\)};
    \vertex (y3) at (6.2,-1.6);
    
    \vertex (textc) at (3.5,-0.25) {\footnotesize \textcolor{red}{\(P_{Vf}\)}};
    
    \diagram* {
      (a) -- [fermion] (b) -- [boson] (f1),
      (b) -- [fermion, edge label'=\small\(\mu^{-}\)] (c),
      (c) -- [fermion] (b2),
      (c) -- [boson, edge label'=\small\(Z\textrm{,}\gamma\)]  (b3) --  (d),
      (b3) -- (y1),
      (b3) -- (y2),
      (b3) -- (y3),
    };
  \end{feynman}
\end{tikzpicture}
\begin{tikzpicture}
  \begin{feynman}
    \vertex (a) at (0.8,0) {\small \(\mu^{-}\)};
    \vertex[dot,red,minimum size=0.15cm] (c) at (3.5,-0.6) {};
    \vertex (b2) at (5,-0.2) {\small \(\mu^{-}\)};
    \vertex[blob,minimum size=0.8cm] (b3) at (5,-1.2) {};
    \vertex (d) at (2.5,-2) {\small \(X\)};
    \vertex (vl1) at (1.6,-0.18);
    \vertex (vl2) at (2.5,-0.36);
    \vertex (y1) at (6.2,-0.8);
    \vertex (y2) at (6.2,-1.2);
    \vertex (nome) at (6.5,-1.2) {\small \(Y\)};
    \vertex (y3) at (6.2,-1.6);

    \vertex (text1) at (2.8,-0.7) {\(\mu^{-}\)};
    \vertex (text2) at (2.2,0.5) {\small \(V\)};
    \vertex (textc) at (3.5,-0.25) {\footnotesize \textcolor{red}{\(P_{Vf}\)}};
    
    \diagram* {
      (a) -- [] (vl1),
      (vl1) --  [fermion] (vl2),
      (vl1) -- [boson, out=90, in=60, looseness=2.] (vl2),
      (vl2) -- [] (c),
      (c) -- [fermion] (b2),
      (c) -- [boson, edge label'=\small\(Z\textrm{,}\gamma\)]  (b3) --  (d),
      (b3) -- (y1),
      (b3) -- (y2),
      (b3) -- (y3),
    };
  \end{feynman}
\end{tikzpicture}
\caption{Sketch of the contribution due to the $P_{Vf}$ splitting at $\mathcal{O}(\alpha^2)$, accounting for a double emission or virtual correction plus one collinear emission.} 
\label{fig:fsplitapp}
\end{figure}
We now discuss the NLO contribution due to the $P_{Vf}$ splitting. The zeroth order $f_{\mu_L}^{(0)}=f_{\mu_R}^{(0)}=\frac{1}{2}\delta (1-x)$ appearing in \cref{eq:appEVA} lead to the EVA approximation of the $Z/\gamma$ PDF that we have already discussed in \cref{sec:NLO}. Therefore, we now focus only on the other terms appearing at $\mathcal{O}(\alpha)$ order in $f_{\mu_{L/R}}$, \textit{i.e.} on the double emission diagrams (or virtual correction plus emission) sketched in \cref{fig:fsplitapp}. These convolutions in \cref{eq:appa2split} are, precisely
\be
\begin{split}
&Q^Z_{\mu_L} P^{f}_{V_+ f_L}\otimes f^{(\alpha)}_{\mu_L} + Q^{Z}_{\mu_R} P^{f}_{V_- f_L} \otimes f^{(\alpha)}_{\mu_R}=\\
&=\Big(-\frac{1}{2}+s_W^2\Big) \int_{x}^1 \,\frac{dz}{z}\,\frac{(1-z)^2}{z}\,f_{\mu_L}^{(\alpha)}\Big(\frac{x}{z},Q^2\Big)\, +\, (0+s_W^2) \int_{x}^1 \,\frac{dz}{z}\,\frac{1}{z}\,f_{\mu_R}^{(\alpha)}\Big(\frac{x}{z},Q^2\Big)\,.
\end{split}
\ee
After performing these integrals and further integrating over the evolution scale, we find
\be\label{eq:fZgamma_Vf}
\begin{split}
f^{(\alpha^2) P_{Vf}}_{Z/\gamma_{+}}(x,Q^2) &= \frac{\alpha_{\gamma 2}\alpha_2}{48 c_W \pi^2}(t-t_Z)^3\frac{(1-x)^2}{x}\Big(-\frac{1}{2}+s_W^2\Big)-\\[2mm]
&-\frac{\alpha_{\gamma 2}\alpha_2}{64 c_W^3 \pi^2 x}(t-t_Z)^2 S(x)-\frac{\alpha_{\gamma 2}\alpha_{\gamma}}{16 c_W \pi^2 x}(t^2-t_Z^2) P(x)\,,
\end{split}
\ee
where $P(x)$ and $S(x)$ explicitely read
\be
\begin{split}
S(x) = &-3 c_W^2 + 6 c_W^2 s_W^2 - x + 6 c_W^2 x + 6 s_W^2 x - 12 c_W^2 s_W^2 x - 12 s_W^4 x +\\
&+16 s_W^6 x + x^2 - 3 c_W^2 x^2 - 6 s_W^2 x^2 + 6 c_W^2 s_W^2 x^2 + 
 12 s_W^4 x^2 - 4 s_W^6 x^2 +\\
&+ 2 (-(-1 + x)^2 + 6 s_W^2 (-1 + x)^2 - 12 s_W^4 (-1 + x)^2 + 8 s_W^6 (2 - 2 x +\\
&+ x^2)) \log(1 - x)-(-1 + 2 s_W^2)^3 (-2 + x) x \log(x)\,,\\[2mm]
P(x) = &\, (-2 (-1 + x)^2 + 4 s_W^2 (2 - 2 x + x^2)) \log(1 - x) + 
 x (-1 - s_W^2 (-4 + x)+\\
& + x - (-1 + 2 s_W^2) (-2 + x) \log(x))\,.
\end{split}
\ee
It can be noted the presence of a double Sudakov logarithm $\alpha^2 (t-t_Z)^3$. It is originated in the virtual correction to the $\mu_L$ PDF due to a $W$ boson loop, as in the right diagram of \cref{fig:fsplitapp}.
\subsection*{$P_{Vh}$ contribution at $\mathcal{O}(\alpha^2)$}
\begin{figure}[th]
\centering
\begin{tikzpicture}
  \begin{feynman}
    \vertex (a) at (0,0) {\small \(\mu^{-}\)};
    \vertex (b) at (2,0);
    \vertex (f1) at (4,0.4) {\small \(\nu_{\mu}\)};
    \vertex[dot,red,minimum size=0.15cm] (c) at (3.5,-0.6) {};
    \vertex (b2) at (5,-0.2) {\small \(W^-_{L}\)};
    \vertex[blob,minimum size=0.8cm] (b3) at (5,-1.2) {};
    \vertex (d) at (2.5,-2) {\small \(X\)};
    \vertex (y1) at (6.2,-0.8);
    \vertex (y2) at (6.2,-1.2);
    \vertex (nome) at (6.5,-1.2) {\small \(Y\)};
    \vertex (y3) at (6.2,-1.6);
    
    \vertex (textc) at (3.5,-0.25) {\footnotesize \textcolor{red}{\(P_{Vh}\)}};
    
    \diagram* {
      (a) -- [fermion] (b) -- [fermion] (f1),
      (b) -- [boson, edge label'=\small\(W^{-}_{L}\)] (c),
      (c) -- [boson] (b2),
      (c) -- [boson, edge label'=\small\(Z\textrm{,}\gamma\)]  (b3) --  (d),
      (b3) -- (y1),
      (b3) -- (y2),
      (b3) -- (y3),
    };
  \end{feynman}
\end{tikzpicture}
\caption{Sketch of the contribution due to the $P_{Vh}$ splitting at $\mathcal{O}(\alpha^2)$.} 
\label{fig:WLsplitapp}
\end{figure}
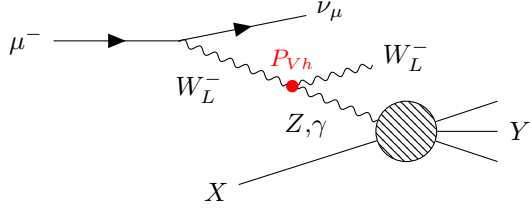
Finally, we discuss the contribution due to the collinear emission of a longitudinally polarized $W^-_L$ boson, sketched in \cref{fig:WLsplitapp}. Explicitely, the convolution in \cref{eq:appa2split} describing this process reads
\be
\begin{split}
P^{h}_{V_+ h}\otimes f_{W^-_L}^{(\alpha)}&=\int_{x}^1\,\frac{dz}{z}\,\frac{(1-z)}{z} f_{W^-_L}^{(\alpha)}\Big(\frac{x}{z},Q^2\Big)=\\
&=\int_{x}^1\,\frac{dz}{z}\,\frac{(1-z)}{z}\frac{(z-x)\alpha_2}{4 \pi x}= -\frac{\alpha_2 }{4 \pi } \frac{(2-2 x+(x+1)\log (x))}{x}\,.
\end{split}
\ee
Then, the integral over the factorization scale $t$ is trivial and amounts to the overall contribution
\be\label{eq:fZgamma_Vh}
\begin{split}
f^{(\alpha^2) P_{Vh}}_{Z/\gamma_{+}}(x,Q^2) =
-\frac{\alpha_2 \alpha_{\gamma 2}}{8 \pi ^2 x\, c_W}\left(c_W^2- s_W^2\right)
   (t-t_Z) (2-2 x+(x+1) \log (x))\,.
\end{split}
\ee
The three contributions described in these sections, added to the EVA approximation of $f_{Z/\gamma}$, were used in \cref{sec:NLO} to plot the $\mathcal{O}(\alpha^2)$ PDFs of \cref{fig:comparison}.

\subsection*{Analytic expressions for $f_ {Z/\gamma_-}(x,Q^2)$}

We report here the analytic contributions at order $\mathcal{O}(\alpha^2)$ to the $f_ {Z/\gamma_-}(x,Q^2)$ PDF. These have been determined following the same procedure outlined above for the positive helicity:
\be
\begin{split}
f^{(\alpha^2) P_{VV}}_{Z/\gamma_{-}}(x,Q) = &\, \frac{\alpha_2 \alpha_{\gamma 2}}{96 \pi^2 x}(t-t_Z)^2 \,c_W \cdot \Big[ 4(t-t_Z) + K(x)\Big]\,,\\[2mm]
f^{(\alpha^2) P_{Vf}}_{Z/\gamma_{-}}(x,Q) = &\,\frac{\alpha_{\gamma 2}\alpha_2}{48 c_W \pi^2}(t-t_Z)^3\frac{1}{x}\Big(-\frac{1}{2}+s_W^2\Big)-\\[2mm]
-&\frac{\alpha_{\gamma2} \alpha_2}{128 c_W^3 \pi^2x}(t-t_Z)^2 R(x)-\frac{\alpha_{\gamma 2}\alpha_{\gamma}}{32 c_W \pi^2 x}(t^2-t_Z^2)T(x)\,,\\[2mm]
f^{(\alpha^2) P_{Vh}}_{Z/\gamma_{-}}(x,Q) = &\,-\frac{\alpha_2 \alpha_{\gamma 2}}{8 \pi ^2 x\, c_W}\left(c_W^2- s_W^2\right)
   (t-t_Z) (2-2 x+(x+1) \log (x))\,,
\end{split}
\ee
where we defined
\be
\begin{split}
K(x)=& -31 - 12 x + 39 x^2 + 4 x^3+12 \log(1 - x) -6 (2 + 6 x + 3 x^2)\log(x)\,,\\[2mm]
R(x)=& -6 c_W^2 + 12 c_W^2 s_W^2 - 2 x + 12 s_W^2 x - 24 s_W^4 x + 
 32 s_W^6 x - x^2 +6 s_W^2 x^2-\\
 & - 12 s_W^4 x^2 - 8 s_W^6 x^2 +4 (-1 + 6 s_W^2 - 12 s_W^4 + 8 s_W^6 (2 - 2 x + x^2))\cdot \\
 &\cdot\log(1 - x) -16 s_W^6 (-2 + x) x \log(x)\,,\\[2mm]
T(x)=&\, 4 (-1 + 2 s_W^2 (2 - 2 x + x^2)) \log(1 - x) -x (2 + 2 s_W^2 (-4 + x)+\\
& + x + 4 s_W^2 (-2 + x) \log(x))\,.
\end{split}
\ee
The Sudakov double logarithms in this formulas originate from the same source as those discussed in the preceding sections. The contribution given by the emission of a longitudinal $W_L$ boson has the same form as in the positive helicity case, since this splitting function is independent of the vector boson's polarization, $P_{V_+ h}(z)=P_{V_- h}(z)$.
\section{Differential cross section}
\label{app:diffxsec}

The differential cross section for a generic partonic process $1\,2 \to 3\,4$ can be written, in the lab frame and in the high energy limit, as:
\begin{equation}\label{eq:TripleDiff}
\frac{d^3 \sigma_{\textrm{tot}}}{dy_3 dy_4 dm} = f_{1}(x_1) f_{2}(x_2) \frac{m^3}{2s}\frac{1}{\textrm{cosh}^2 y_{*}}\frac{d\sigma_{\textrm{H}}}{d t}(1\,2 \to 3\,4) \,.
\end{equation}
where $f_i(x_i)$ are the PDFs, $\sigma_{H}$ is the hard cross section, $m$ is the energy in the center of mass frame, $y_{i}$ is the rapidity of the state $i$ and we defined
\be
x_{1,2}=\frac{m}{\sqrt{s}}e^{\pm \frac{y_3 + y_4}{2}}\,,\quad y_{*}=\frac{1}{2}(y_3-y_4)\,,\quad t=-\frac{m^2}{2}(1-\textrm{cos}\,\theta_{*})\,,\quad \theta_{*}=\textrm{arcsin}\left(\frac{1}{\textrm{cosh}\,{y_{*}}}\right)\,.
\ee
The relation $m=4\,p_T^2\,\textrm{cosh}^2y_*$ allows to express the total cross section as a function of the transverse momentum $p_T$:
\begin{equation}
\frac{d^3 \sigma_{\textrm{tot}}}{dy_3 dy_4 dp_T} = f_{1}(x_1) f_{2}(x_2) \frac{2 m \,p_T }{s}\frac{1}{\textrm{cosh}^2 y_{*}}\frac{d\sigma_{\textrm{H}}}{d t}(1\,2 \to 3\,4) \,.
\end{equation}
A step by step derivation of the above formulas can be found in~\cite{Peskin:1995ev}.

\subsection{Partonic cross sections for Compton scattering}
\label{app:xsecCompton}

In the following, we report the helicity-dependent partonic hard cross section for Compton scattering, on which the results of \cref{sec:Compton} are based:

\be\begin{split}
    \frac{d\sigma}{dt}(\mu_{L,R}^- \gamma_{\mp} \to \mu^- \gamma) &= \frac{4 \pi \alpha^2 (s+t)}{s^3}~, \\
    \frac{d\sigma}{dt}(\mu_{L,R}^- \gamma_{\pm} \to \mu^- \gamma) &= \frac{4 \pi \alpha^2}{s (s + t)}~,
\end{split}\ee

\be\begin{split}
    \frac{d\sigma}{dt}(\mu_{L}^- Z_{-} \to \mu^- \gamma) &=
        \frac{\pi \alpha^2  \left( 1 - 2 s_W^2 \right)^2}{ c_W^2 s_W^2 } 
        \frac{s \left(s + t  -m_Z^2\right)}{\left( s - m_Z^2 \right)^4}~, \\
    \frac{d\sigma}{dt}(\mu_{L}^- Z_{+} \to \mu^- \gamma) &= 
        \frac{\pi \alpha^2  \left( 1 - 2 s_W^2 \right)^2}{ c_W^2  s_W^2  } 
        \frac{\left( t^2 m_Z^4 + \left( s - m_Z^2 \right)^4 \right)}{s \left( s - m_Z^2 \right)^4 \left( s + t - m_Z^2 \right)}~, \\
    \frac{d\sigma}{dt}(\mu_{L}^- Z_{L} \to \mu^- \gamma) &= 
        -\frac{ 2 \pi \alpha^2  \left( 1 - 2 s_W^2 \right)^2}{ c_W^2  s_W^2 }
        \frac{ m_Z^2 t }{s^2 \left( s - m_Z^2 \right)^4}~, \\
    \frac{d\sigma}{dt}(\mu_{R}^- Z_{-} \to \mu^- \gamma) &= 
        \frac{4 \pi \alpha^2  s_W^2 }{ c_W^2 } 
        \frac{\left( t^2 m_Z^4 + \left( s - m_Z^2 \right)^4 \right)}{s \left( s - m_Z^2 \right)^4 \left( s + t -m_Z^2 \right) } ~, \\
    \frac{d\sigma}{dt}(\mu_{R}^- Z_{+} \to \mu^- \gamma) &= 
        \frac{4 \pi \alpha^2  s_W^2}{ c_W^2} 
        \frac{s \left( s + t -m_Z^2 \right)}{\left( s - m_Z^2 \right)^4} ~, \\
    \frac{d\sigma}{dt}(\mu_{R}^- Z_{L} \to \mu^- \gamma) &= 
        -\frac{8 \pi \alpha^2 s_W^2}{ c_W^2} 
        \frac{m_Z^2 t }{\left( s - m_Z^2 \right)^4}~,
\end{split}\ee

\be\begin{split}
    \frac{d\sigma}{dt}(\mu_{L}^- Z/\gamma_{-} \to \mu^- \gamma) &= 
        - \frac{4 \pi \alpha^2 \left( 1 - 2 s_W^2 \right)}{c_W s_W} 
        \frac{s \left( s + t - m_Z^2 \right)}{ \left( s- m_Z^2 \right)^4} ~, \\
    \frac{d\sigma}{dt}(\mu_{L}^- Z/\gamma_{+} \to \mu^- \gamma) &= 
        - \frac{4 \pi \alpha^2 \left( 1 - 2 s_W^2 \right)}{c_W s_W} 
        \frac{\left( t^2 m_Z^4 + \left( s - m_Z^2 \right)^4 \right)}{s \left( s - m_Z^2 \right)^4 \left( s + t - m_Z^2 \right)}~, \\
    \frac{d\sigma}{dt}(\mu_{R}^- Z/\gamma_{-} \to \mu^- \gamma) &=  
        \frac{8 \pi \alpha^2 s_W}{c_W}
        \frac{\left( t^2 m_Z^4 + \left( s - m_Z^2 \right)^4 \right)}{s \left( s - m_Z^2 \right)^4 \left( s + t - m_Z^2 \right)}~, \\
    \frac{d\sigma}{dt}(\mu_{R}^- Z/\gamma_{+} \to \mu^- \gamma) &= 
        \frac{8 \pi \alpha^2 s_W}{c_W}
        \frac{s \left( s + t - m_Z^2 \right)}{ \left( s- m_Z^2 \right)^4} ~.
\end{split}\ee


\bibliographystyle{JHEP}
\bibliography{biblio}

\end{document}